\documentclass[12pt,a4paper,dvips,epsfig,openany]{article}
\usepackage{a4p,rotating}
\usepackage{epsfig,subfigure}
\usepackage{cite,mcite}
\usepackage{graphicx}
\usepackage{physics}
\usepackage{l3_titlePH,ifthen,Lep}
\journalname{Phys. Lett. B}
\date{June 14, 2004}
\preprint{2004-026}
\graphicspath{{/l3/paper/example/}}
\newlength{\capindent}
\newlength{\capwidth}
\newlength{\figwidth}
\newcommand{\icaption}[2][!*!,!]{\hspace*{\capindent}%
  \begin{minipage}{\capwidth}
    \ifthenelse{\equal{#1}{!*!,!}}%
      {\caption{#2}}%
      {\caption[#1]{#2}}
  \end{minipage}}
\input{l3aty2k.def}

\begin{document}
\begin{titlepage}
\title{{\LARGE
    Measurement of the Cross Section of \\
    W-boson pair production at LEP \\
}}

\author{The L3 Collaboration}

%
%
\vspace{-0.5cm}
\begin{abstract}
  The cross section of W-boson pair-production is measured with the L3
  detector at LEP.  In a data sample corresponding to a total
  luminosity of 629.4~pb$^{-1}$, collected at centre-of-mass
  energies ranging from 189 to 209$~\GeV$, 9834
  four-fermion events with W bosons decaying into hadrons or leptons
  are selected.
  The total cross section is measured with a precision of 1.4\% and agrees with the
  Standard Model expectation.
  Assuming charged-lepton universality, the branching fraction for hadronic W-boson decays is
  measured to be: $ \mathrm{Br}(\mathrm{W\rightarrow hadrons}) = 67.50 \pm
  0.42~(\mathrm{stat.}) \pm 0.30~(\mathrm{syst.})~\%$, in agreement with the Standard
  Model. Differential cross sections as a
  function of the W$^-$ production angle are also measured for the
  semi-leptonic channels $\QQEN$ and $\QQMN$.
\end{abstract}
%
%
\submitted

\end{titlepage}
%
%
\section{Introduction}

From 1996 until the year 2000, the centre-of-mass energy,
$\sqrt{s}$, of the LEP $\EE$ collider at CERN was increased
in several steps from $161~\GeV$ to $209~\GeV$.
These energies, being above the kinematic threshold of
W-boson pair production, allow 
detailed studies of this process. 

\begin{figure}[p]
\begin{center}
\epsfig{file=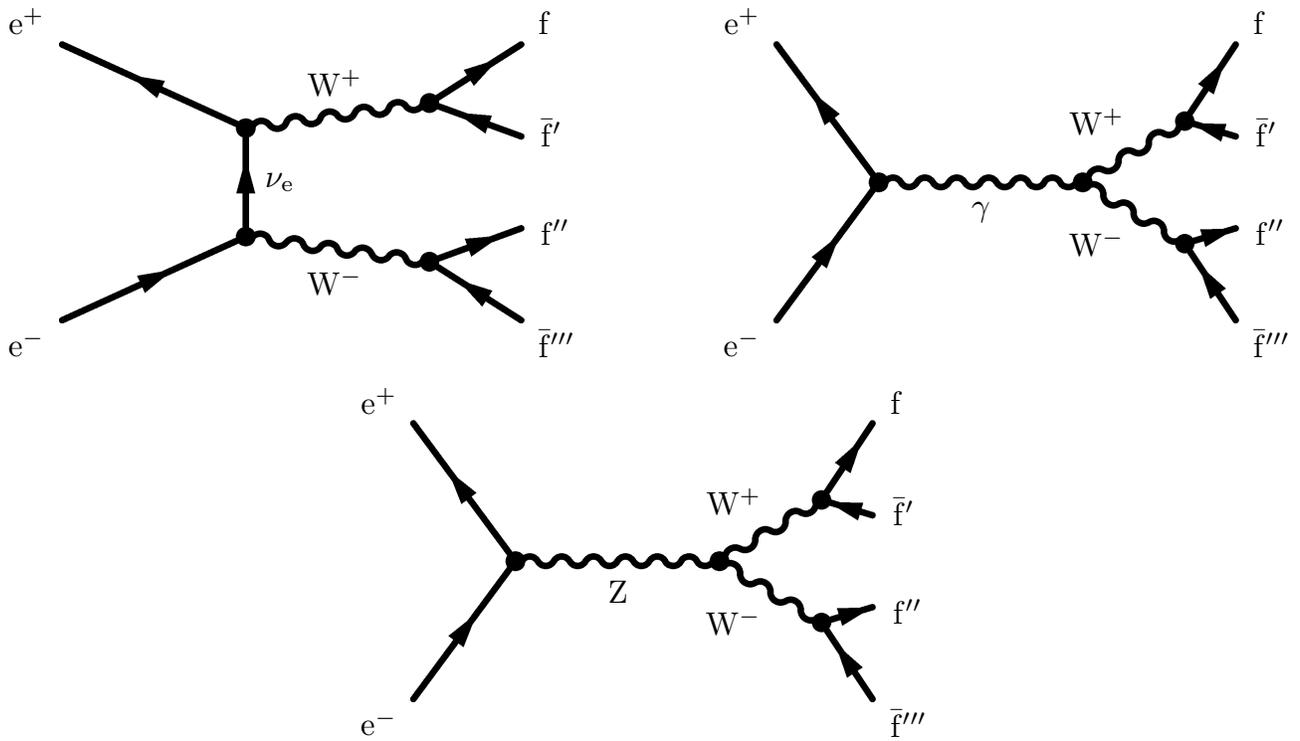,width=\linewidth}
\vskip -0.1cm
\caption[]{The lowest-order Feynman diagrams (CC03) contributing to
  W-boson pair
  production: $t$-channel $\nu$ exchange 
  and $s$-channel $\gamma$ and Z-boson exchange.} 
\label{fig:cc03}
\end{center} 
\end{figure}

To lowest order within the Standard Model~\cite{standard_model}, three
 charged-current Feynman diagrams,  shown in Figure~\ref{fig:cc03} and
 referred to as CC03~\cite{CCNC,LEP2YRWW,LEP2YREG}, 
yield four-fermion final states  via W-boson pair production: 
 $t$-channel $\nu$ exchange and $s$-channel
$\gamma$ and Z-boson exchange.
W bosons decay
into a quark-antiquark pair
or a lepton-antilepton pair, denoted here as
$\mathrm{qq}$, $\ell\nu$ ($\ell=\e,\mu,\tau$) or, in general, $\FF$ for both
W$^+$ and W$^-$ decays.  
This Letter describes measurements of all four-fermion final states
 $\LNLN$, $\QQLN$ and $\QQQQ$ mediated
by W-boson pair production.
The presence of additional photons in the final state is not excluded.
Contributions to the production of four-fermion final
states arising from other neutral- or charged-current 
Feynman diagrams are small.  
At the current level of statistical
accuracy, interference terms are sizable 
for the 151 charged- and neutral-current diagrams contributing
to the $\LNLN$ final states, for the 20 charged-current diagrams
contributing to the $\QQEN$ final state and for the 214 charged- and
neutral-current diagrams contributing
to the $\QQQQ$ final state\cite{CCNC,LEP2YRWW,LEP2YREG}.

It is conventional to quote results for the CC03 subset of diagrams,
including the effect of initial-state radiation. As
four-fermion states produced by all diagrams are measured, a suitable reweighting
technique, described in the following, is applied to extract these results.

W-boson decay branching fractions and 
the total W-boson pair-production cross section are determined with improved
precision as compared to earlier L3 measurements at
$\sqrt{s}=161-189\,\GeV$~\cite{l3-111,l3-120,l3-155,l3-218}. Comparable
results were reported by other LEP
experiments~\cite{ADO-XSEC-ALL}. 
 
The differential cross sections for the $\QQEN$ and 
$\QQMN$ final states, as a function of the W$^-$ production
angle with respect to the direction of the incoming electron, are also derived.

%
\section{Data and Monte Carlo Samples}
\label{sec:samples}

The results presented  in this Letter are based on the full luminosity
collected by the L3 detector~\cite{l3-01-new} during the high-energy runs of the LEP
collider. The measurement of the total luminosity, ${\cal L}$, follows the
procedure described in Reference~\citen{l3-lumi96}.  

The data collected at $\sqrt{s}=192-209\,\GeV$ are analysed in seven
$\sqrt{s}$ bins,  as detailed in Table~\ref{tab:stat}.
The corresponding centre-of-mass energies are known with a precision
of about $50~\MeV$~\cite{LEPECAL00}.
Results based on data collected at $\sqrt{s}=189\,\GeV$ were already 
published~\cite{l3-218} but are reanalysed here since
improved Monte Carlo programs are now available for signal simulation:
KandY~\cite{KandY} and RacoonWW~\cite{RACOONWW}. 
The KandY generator combines the four-fermion generator KORALW~\cite{KORALW}
with the $\cal{O}(\alpha)$ corrections to W-boson pair production 
as implemented in the YFSWW3~\cite{YFSWW3} program. These corrections
are calculated using the leading-pole approximation~\cite{DPA}. The
RacoonWW Monte Carlo program implements such corrections in the
double-pole approximation with similar accuracy and it is used 
for the estimation of systematic uncertainties. 

All KandY Monte Carlo samples used in this analysis are
generated using the full set of Feynman diagrams contributing to a
specific four-fermion final state. 
The KandY program provides the matrix elements on
an event-by-event basis for different contributions including, for example, 
the CC03 subset of diagrams or the ${\cal O}(\alpha)$ corrections. This feature is
extensively used in the following, both to derive quantities at CC03
level and for the assessment  of systematic uncertainties. 
For example, the CC03-level efficiencies are calculated by reweighting every event with the
factor $w_\mathrm{CC03} = |{\cal M}(\mathrm{CC03})|^2/|{\cal M}(\mathrm{4f})|^2$, where
${\cal M}(\mathrm{CC03})$ and ${\cal M}(\mathrm{4f})$ are the matrix
elements of the CC03 subset of diagrams and of the full set, respectively. 
The same events, reweighted by the factor $1-w_\mathrm{CC03}$,
describe the remaining four-fermion contribution not arising
from W-boson pair production. In the following they are considered
as background.
As a cross-check, selection
efficiencies are also derived using the 
EXCALIBUR~\cite{EXCALIBUR} four-fermion generator.

The following Monte Carlo
event generators are used to simulate the
background processes: 
KK2f~\cite{KK2f},
PYTHIA~\cite{PYTHIA},
BHAGENE3~\cite{BHAGENE} and 
BHWIDE~\cite{BHWIDE}  for fermion-pair  production, denoted as $\EE
\rightarrow \mathrm{f \kern 0.1em \bar f}(\gamma)$; TEEGG~\cite{TEEGG}
for radiative $\EE\rightarrow \EE \gamma (\gamma)$ events;
DIAG36~\cite{DIAG36} and LEP4F~\cite{LEP4F} for two-photon
collisions with lepton-pair final states and  PHOJET~\cite{PHOJET} for
two-photon collisions with hadronic final states.

Quark fragmentation and hadronisation processes 
are simulated using PYTHIA. Its parameters are
tuned to describe  hadronic Z decays at
$\sqrt{s}=91~\GeV$~\cite{tablepaper}. A dedicated parameter
set, derived from a light-quark Z-decay data sample, 
is used for the W-boson pair-production simulations. Bose-Einstein correlations between 
hadrons from W decays are simulated using the LUBOEI
BE$_{32}$ model~\cite{LUBOEI}, with Bose-Einstein correlations
only between hadrons originating from the same W boson, as supported by our study~\cite{l3-257}. 

The response of the L3 detector is modelled with the
GEANT~\cite{xsigel} detector simulation program which includes the
effects of energy loss, multiple scattering and showering in the
detector material.  Hadronic showers are simulated with the
GHEISHA~\cite{xgheisha} program. Time-dependent detector
inefficiencies, as monitored during data taking, are included in the
simulations.

\section{Four-Fermion Event Selection}
\label{sec:selections}

The selections of four-fermion final states
are designed to mimimise the uncertainty on the cross section
of each channel. They are chosen to be mutually
exclusive, by using complementary cuts, in order to avoid double counting
of events.

Electrons are identified as energy depositions
in the BGO electromagnetic calorimeter having an electromagnetic shower
shape and matching in azimuth a track reconstructed in the
central tracking chamber.  Muons are identified as tracks
reconstructed in the muon chambers, which point back to the interaction
vertex. Tracks which match a minimum-ionising-particle  signature in the
calorimeters are also retained as muon candidates and denoted as MIPs.  Jets
arising from hadronic tau decays are reconstructed using a
jet-clustering algorithm in a cone of $15^\circ$ half-opening
angle~\cite{GEOMJETS}. The momentum of the
neutrino in $\QQLN$ events is identified with the missing momentum
vector of the event. Hadronic jets corresponding to quarks are
reconstructed using the Durham jet algorithm~\cite{DURHAM}. In the
$\EEQQLN$ selections, the hadronic jets are formed from
energy depositions and tracks not belonging to the reconstructed lepton.

Efficiencies are evaluated for each $\sqrt{s}$ point in the form of 10 by
10 matrices  relating events
at CC03 level to those at reconstruction level. 
An example is given in Table~\ref{tab:xmat-1} for $\sqrt{s} =
206.5~\GeV$. Selection efficiencies
at other centre-of-mass energies are only marginally different.

The number of selected events and background contributions
are detailed in Table~\ref{tab:xsec1}. 
A more detailed description of all 
selections is given below.

\vspace{0.5cm}

\subsection{The \boldmath\protect$\LNLN$ Selection}
\label{sec:lnln}

The event selection for the process $\EELNLN$ requires two charged
leptons and missing energy due to the neutrinos.  The selection
depends on whether the event contains zero, one or two identified
electrons or muons, referred to as jet-jet, lepton-jet and
lepton-lepton classes.  
For the lepton-jet and jet-jet classes, only the most energetic jets
are retained as tau candidates.
Electrons, muons and jets from hadronic tau decays are identified
within the polar angular range $|\cos\theta|<0.96$, where $\theta$
is the lepton angle with respect to the beam direction.  For events
with one or two electrons, one electron is required to satisfy
$|\cos\theta|<0.92$ in order to reduce the background
from Bhabha scattering.  For the jet-jet class, the two
most energetic jets must also satisfy $|\cos\theta|<0.92$.

The acoplanarity, defined as the complement of the 
angle between the directions of the two leptons in the plane
transverse to the beam direction, must be greater than 8 degrees
for the lepton-lepton and lepton-jet classes and
14 degrees for the jet-jet class. These criteria
suppress the dominating backgrounds from lepton-pair production and cosmic rays.
The leptons must have a signal in the scintillator time-of-flight counters
compatible with the beam crossing. 
The total momentum transverse to the beam direction, $P_\mathrm{t}$, must
be greater than 8 GeV. 

Events belonging to the lepton-lepton class are selected by requiring an energy
of at least $25~\GeV$ for the more energetic lepton and $5~\GeV$
for the less energetic one. 
For the  lepton-jet class, the energies of the lepton
and of the jet must exceed $20~\GeV$ and $8~\GeV$, respectively. 
For the jet-jet class, the energies of the most energetic and second most
energetic jets must be greater than  $20~\GeV$ and $6~\GeV$,
respectively. 

\begin{figure}[p]
\begin{center}
\begin{tabular}{cc}
{\epsfig{file=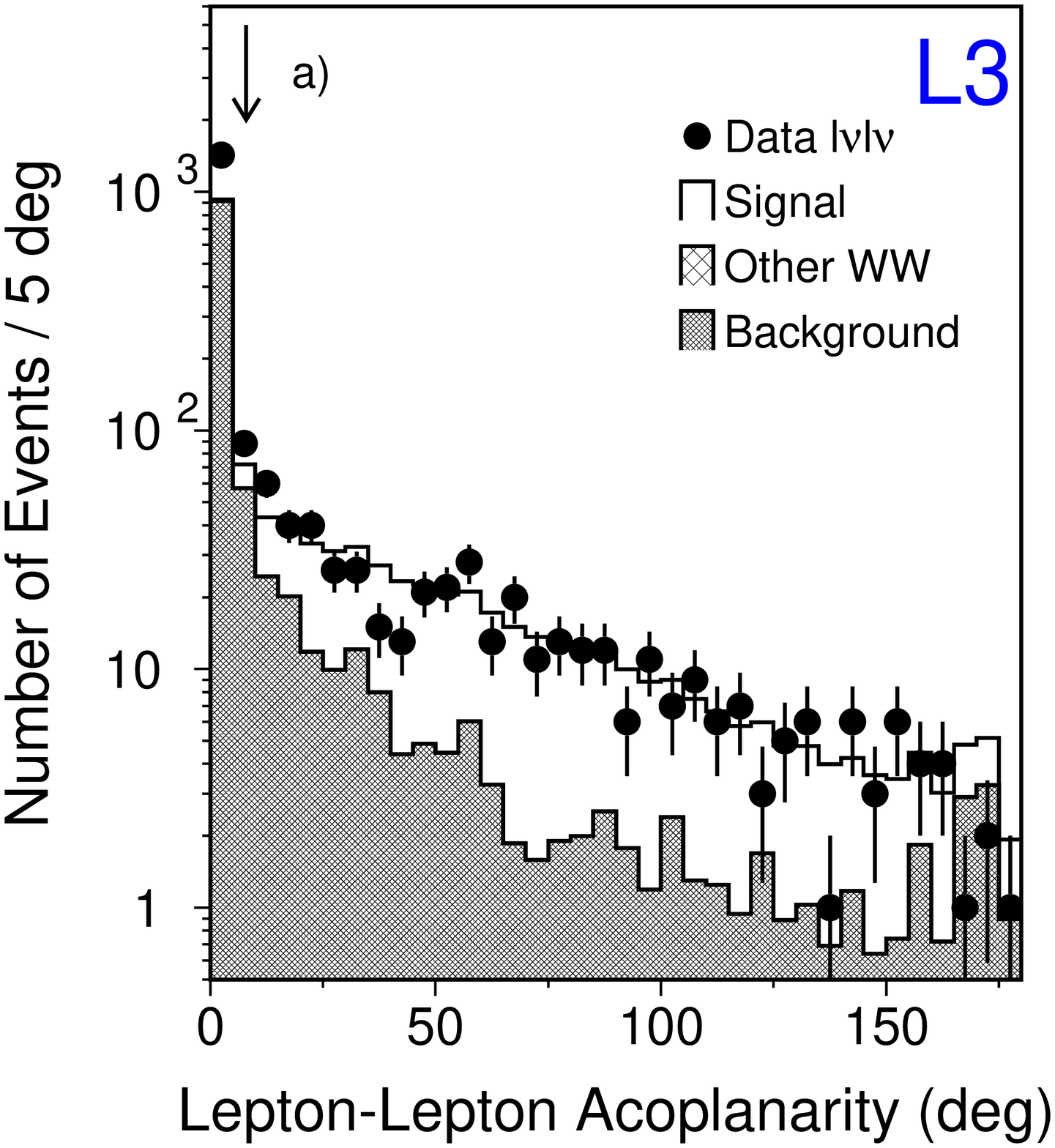,bb=10 30 560 650,width=0.5\linewidth}} &
{\epsfig{file=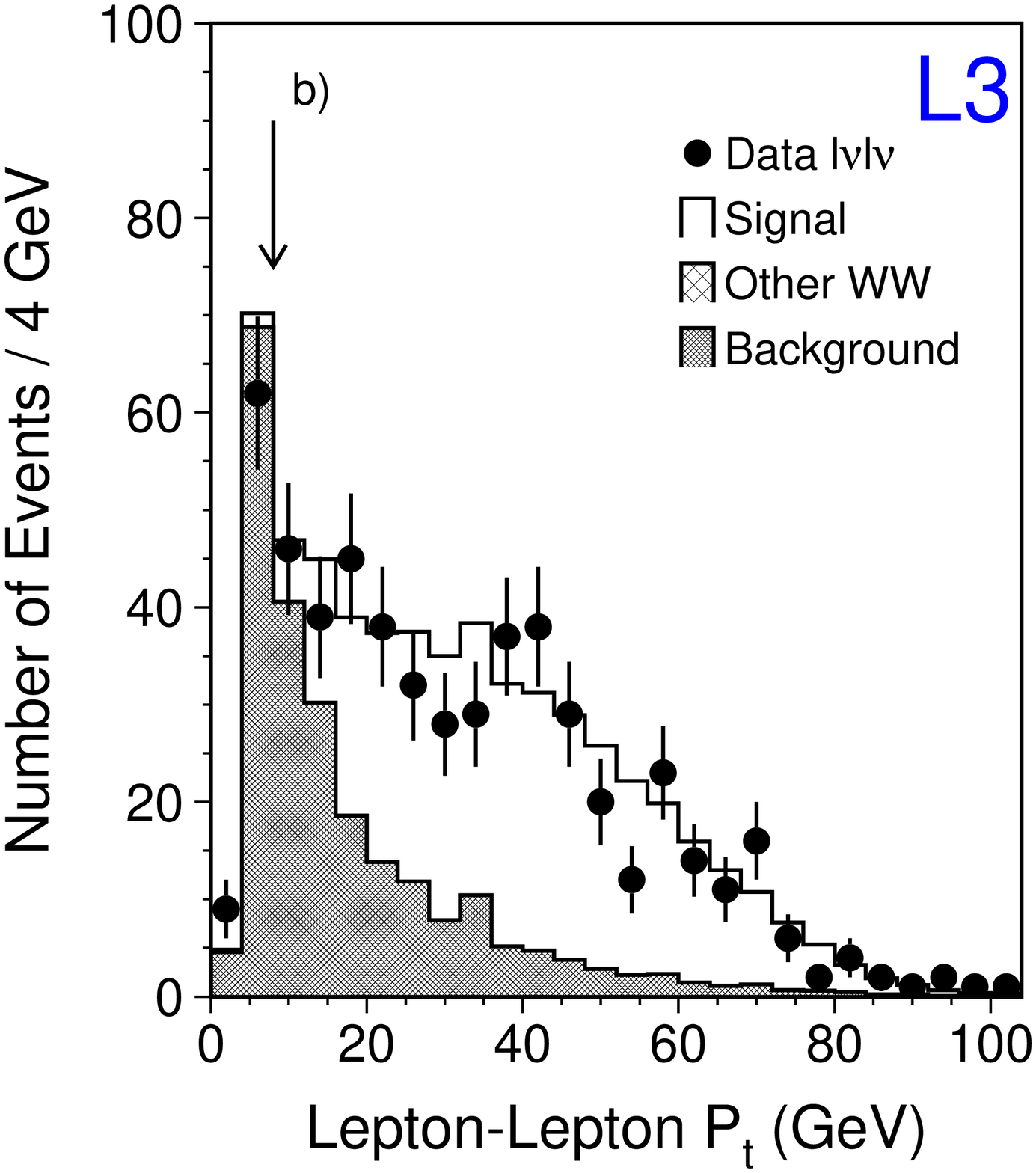,bb=10 30 560 650,width=0.5\linewidth}} \\
{\epsfig{file=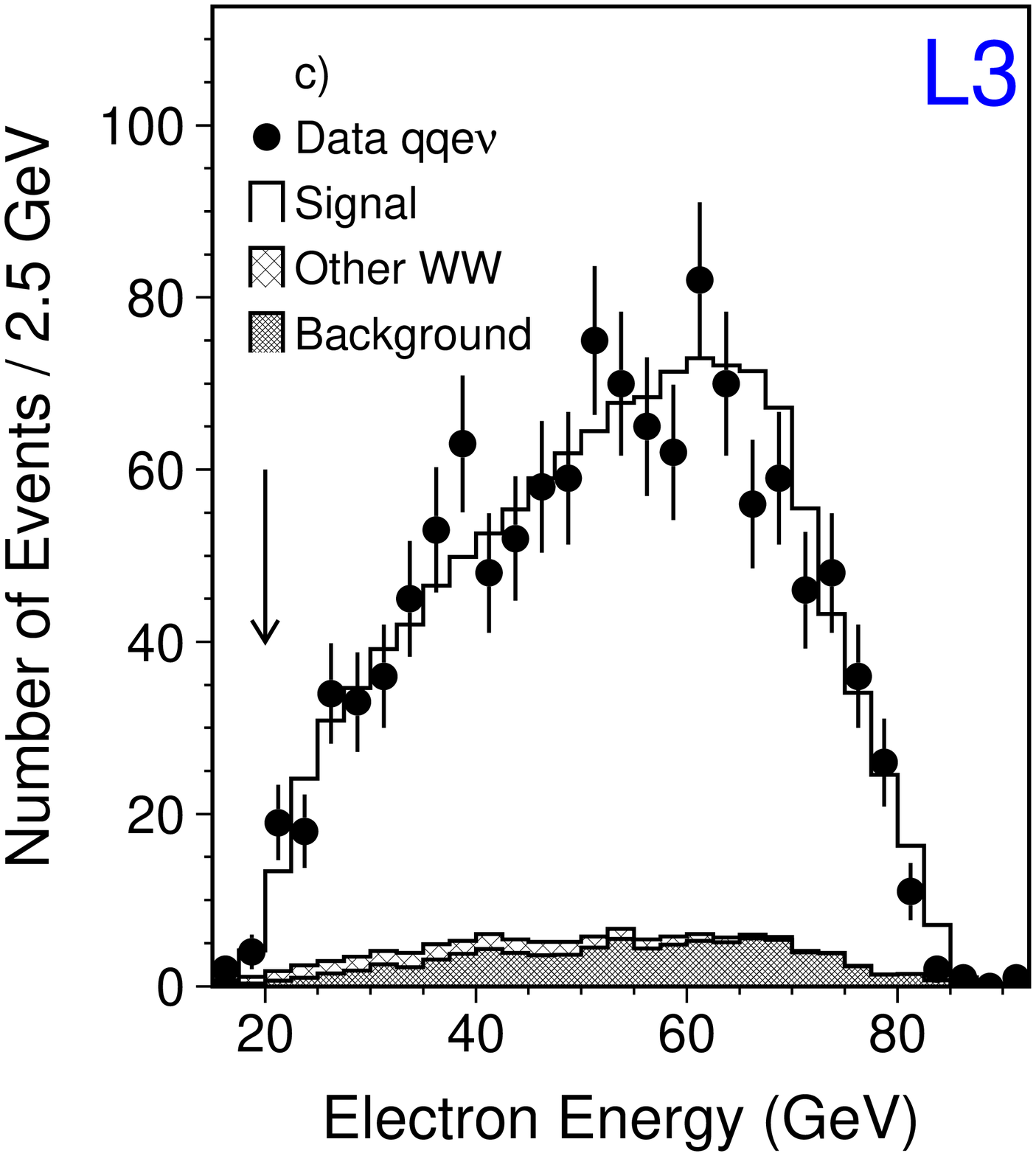,bb=10 30 560 650,width=0.5\linewidth}} &
{\epsfig{file=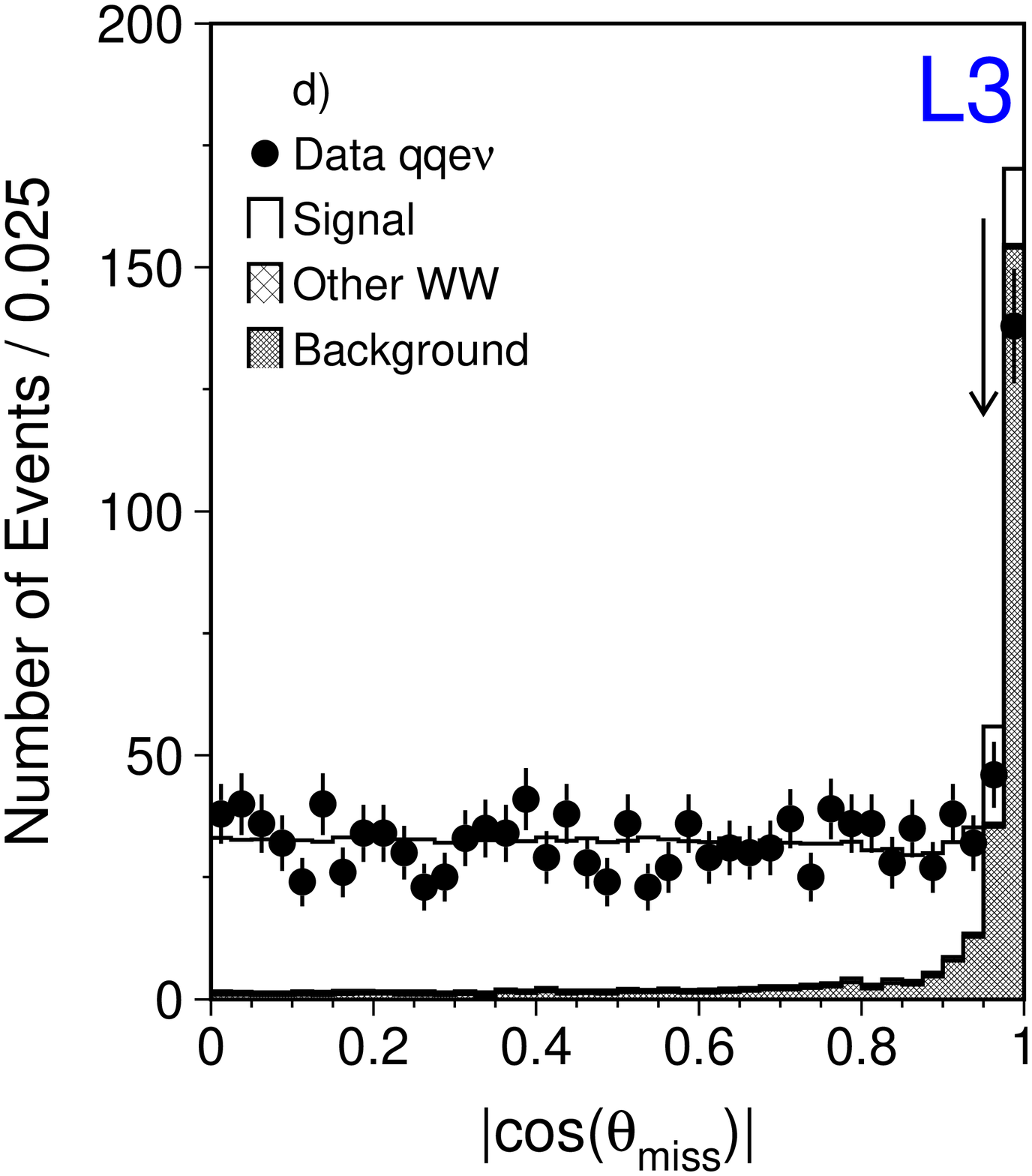,bb=10 30 560 650,width=0.5\linewidth}} \\
\end{tabular}
\caption[]{Distributions of variables used for the selection of
  $\LNLN$ and $\QQEN$ events, 
  comparing the signal and background Monte Carlo to the
  data collected at $\sqrt{s}=189-209\, \GeV$. 
  The positions of the selection cuts are indicated by vertical
  arrows.  All selection cuts except the one on the plotted variable 
  are applied.
  (a) The acoplanarity between the two leptons in the lepton-lepton
  class of the $\LNLN$ selection.
  The excess in the first bin is due to cosmic-ray background.  
  (b) The momentum transverse to the beam
  direction of the selected $\LNLN$ events in the lepton-lepton class.
  (c) The energy of the electron in $\QQEN$ events.
  (d) The absolute value of the cosine of the polar angle of the
  missing momentum in $\QQEN$ events.

}
\label{fig:lnlnqqen}
\end{center} 
\end{figure}

The selected sample has a purity of 72\% at $\sqrt{s}=206.5\, \GeV$. 
The remaining background is dominated by lepton production in  two-photon collisions
(50\%) and lepton-pair production (24\%).
The distributions of the acoplanarity and of the missing momentum 
transverse to the beam direction
for the lepton-lepton class are shown in
Figure~\ref{fig:lnlnqqen}.

\subsection{The \boldmath\protect$\QQEN$ Selection}
\label{sec:qqen}

The event selection for the process $\EEQQEN$ requires an identified
electron of at least $20~\GeV$, high particle-multiplicity 
and large missing momentum.

The missing momentum direction must point well inside the detector
with a polar angle, $\theta_\mathrm{miss}$, such that $|\cos \theta_\mathrm{miss}| <
0.95$. 
The reconstructed jet-jet and lepton-neutrino masses, referred to as
$M_\mathrm{jj}$ and $M_{\mathrm{e}\nu}$,  must 
be greater than $45~\GeV$ and $63~\GeV$, respectively. The latter cut
is used to discriminate between $\EEQQEN$ and $\EEQQTN$ events with 
$\tau \rightarrow \mathrm{e}\nu \nu$.
To further suppress the dominant background from 
the $\EEQQG$ process, which is planar, the directions of the electron and of the two jets are required to
subtend  a solid angle of less than 5.3~sr.

The purity of the selection is 98\% at $\sqrt{s}=206.5\, \GeV$. The
accepted background not originating from W-boson pair production
is dominated by 
$\EEQQEN$ final states 
(71\%) and $\EEQQG$ events (29\%).
The distributions of the energy of the electron and of
$|\cos(\theta_\mathrm{miss})|$ 
are shown in Figure~\ref{fig:lnlnqqen}.

\subsection{The \boldmath\protect$\QQMN$ Selection }
\label{sec:qqmn}

The event selection for the process $\EEQQMN$ requires high
particle-multiplicity, an identified muon or a MIP, and
large missing momentum.  

The jet-jet mass must satisfy $25~\GeV<M_\mathrm{jj}<125~\GeV$ for
events with identified muons and $50~\GeV<M_\mathrm{jj}<98~\GeV$ 
for events with MIPs.
The muon-neutrino reconstructed 
mass, $M_{\mu\nu}$, is used as a discriminant against $\EEQQTN$ events with 
$\tau \rightarrow \mu\nu_\tau \nu_\mu$. Its value is required to
exceed $53~\GeV$. 
This cut is not applied for events containing MIPs.  

The discrimination against $\EEQQTN$ events is further enhanced by
requiring the variable $P^\star = | p_\mu | - 10\;\GeV
(\cos\theta^\star +1)$, 
where $p_\mu$ is the momentum of the
muon and $\theta^\star$ is the
decay angle of the muon in the reconstructed W-boson rest frame, to satisfy $P^\star > 18.5~\GeV$.
This requirement is
loosened to $P^\star > 15~\GeV$ for events with MIPs.

The $\EEQQG$ process is a potentially large source of background. It
is reduced by exploiting the fact that it originates muons close to the jets
and that the total missing momentum, if any, points towards the beam direction.
 The product of
$\psi_{\mu j}$, the angle between the muon and the closest jet, and
$\sin \theta_\mathrm{miss}$ is required to be greater than $5.5$
degrees for events with muons  and
greater than $20$ degrees for events with MIPs.

Background due to  $\QQ\mu^+\mu^-$ final states from Z-boson pair production in events
containing MIPs is rejected by requiring the
relativistic velocity  of the reconstructed W
bosons to be greater than a $\sqrt{s}$-dependent value, ranging from
$0.34$ to $0.49$.

\begin{figure}[p]
\begin{center}
\begin{tabular}{cc}
{\epsfig{file=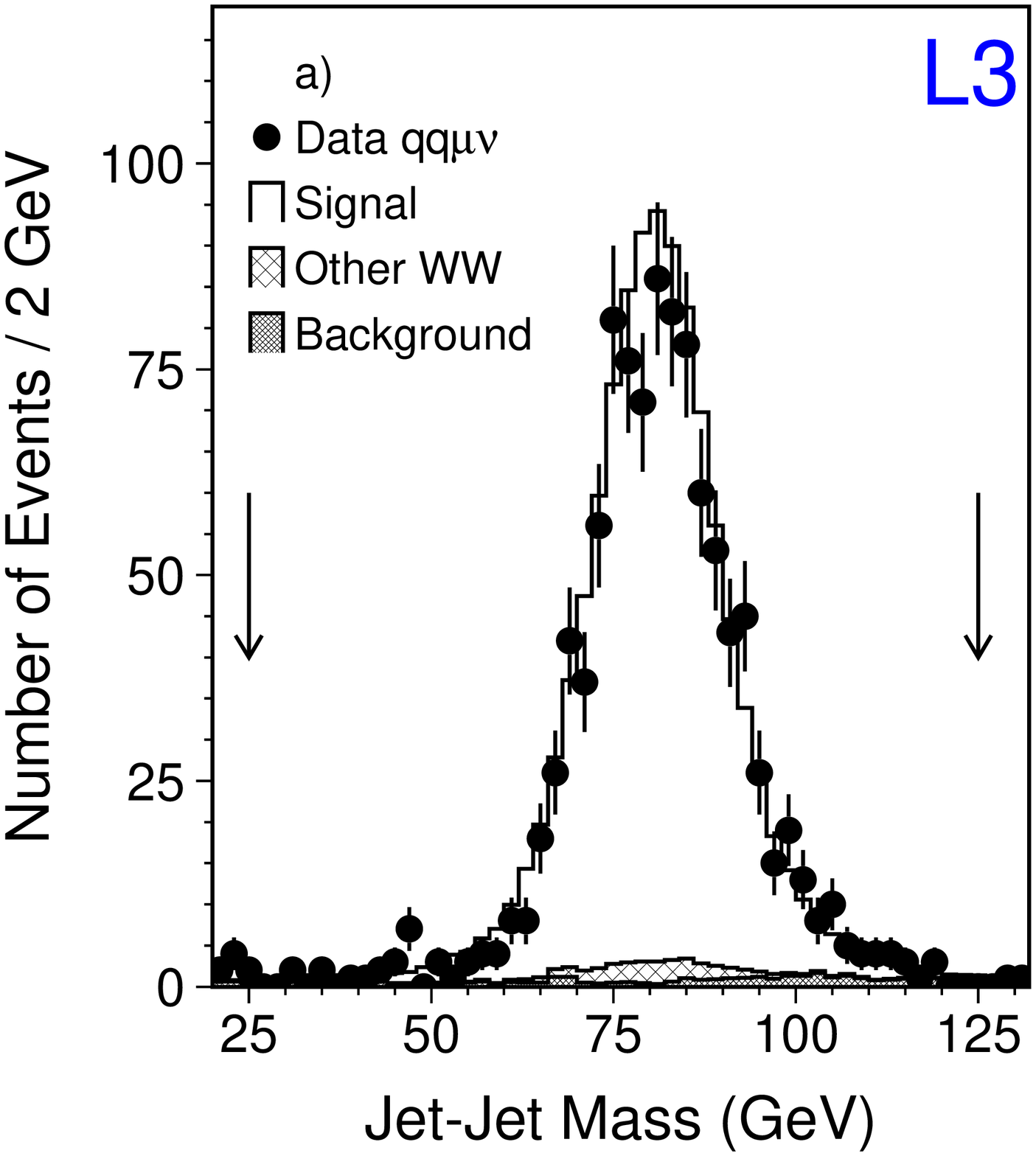,bb=10 30 560 650,width=0.5\linewidth}} &
{\epsfig{file=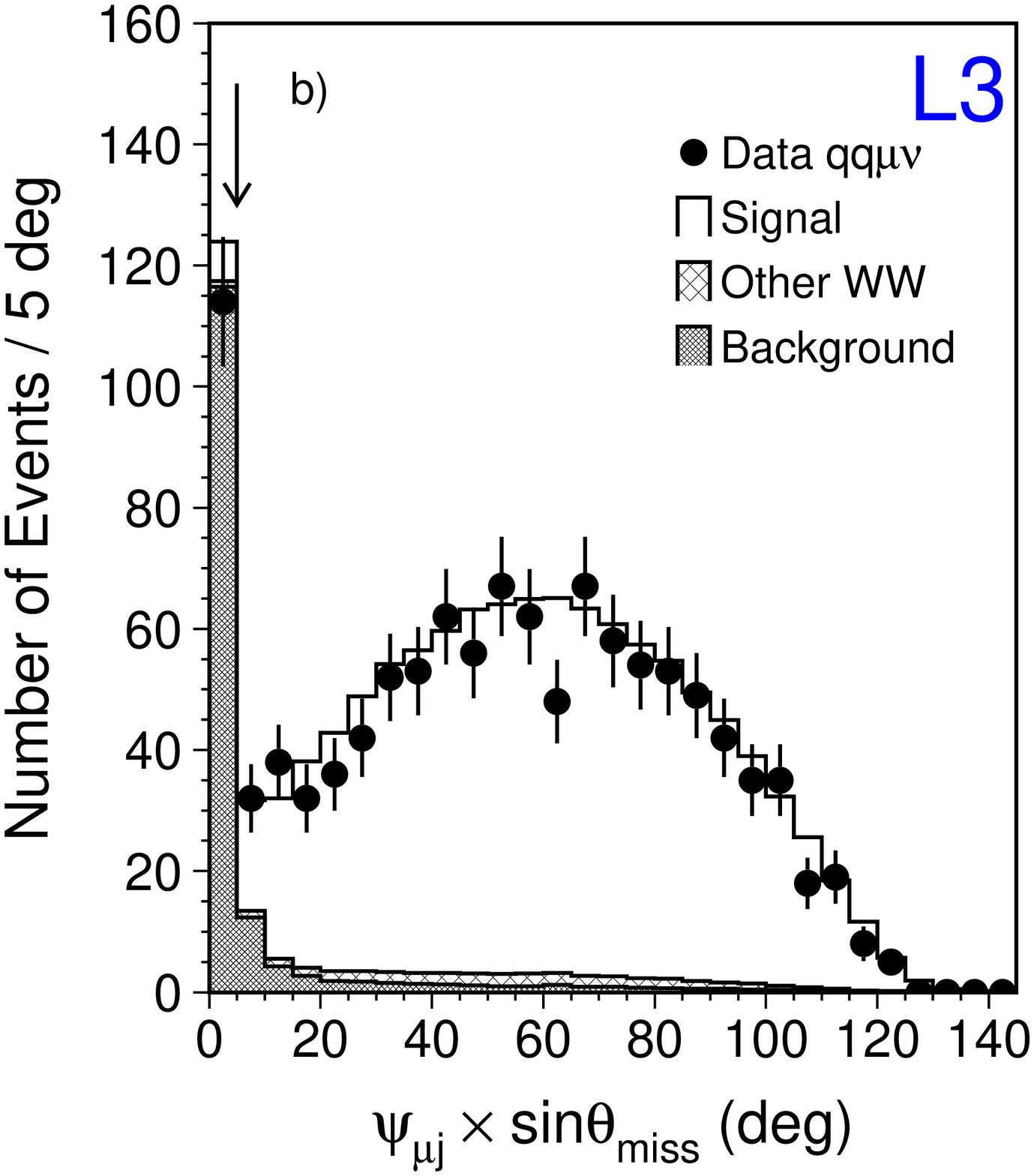,bb=10 30 560 650,width=0.5\linewidth}} \\
{\epsfig{file=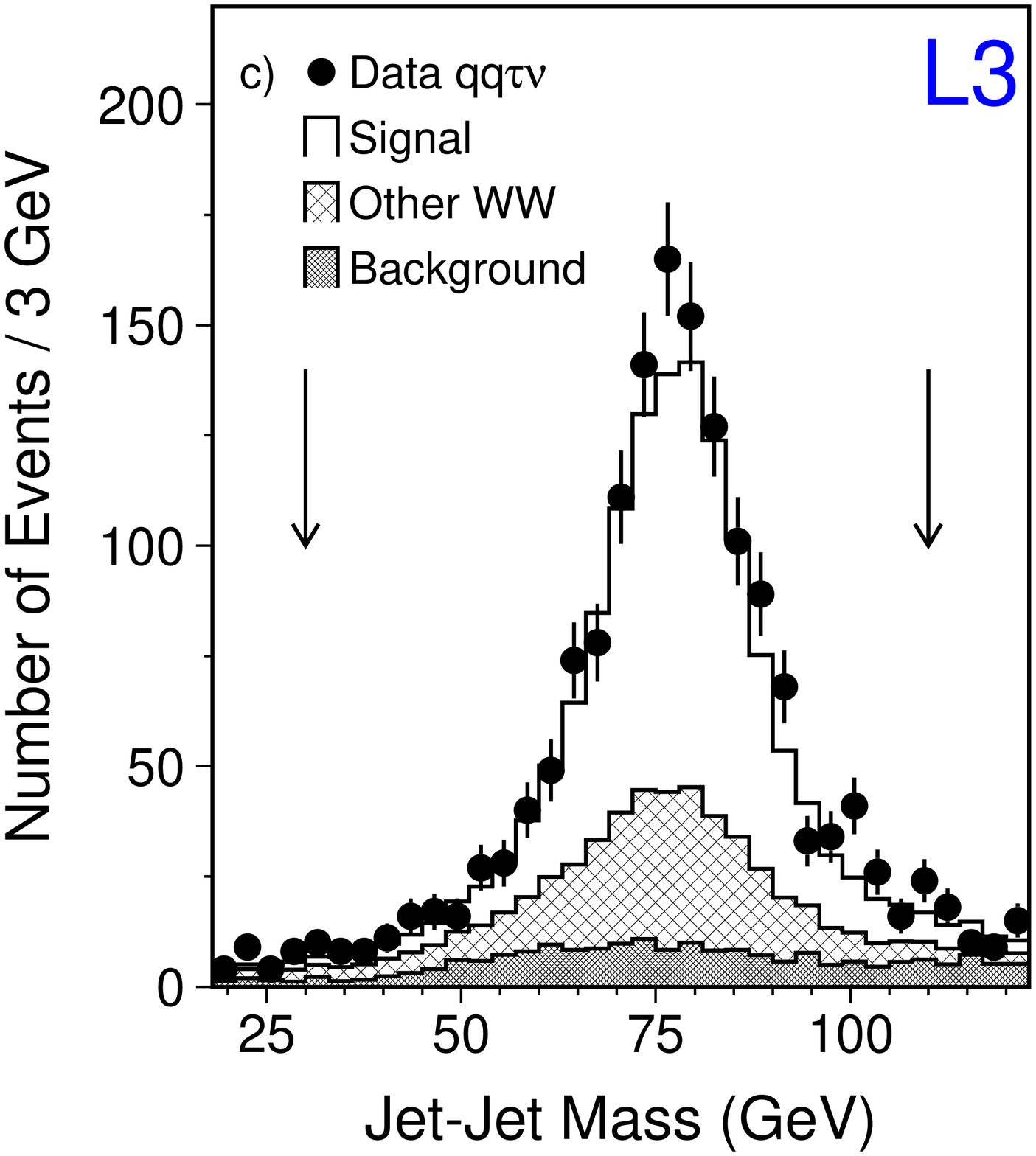,bb=10 30 560 650,width=0.5\linewidth}} &
{\epsfig{file=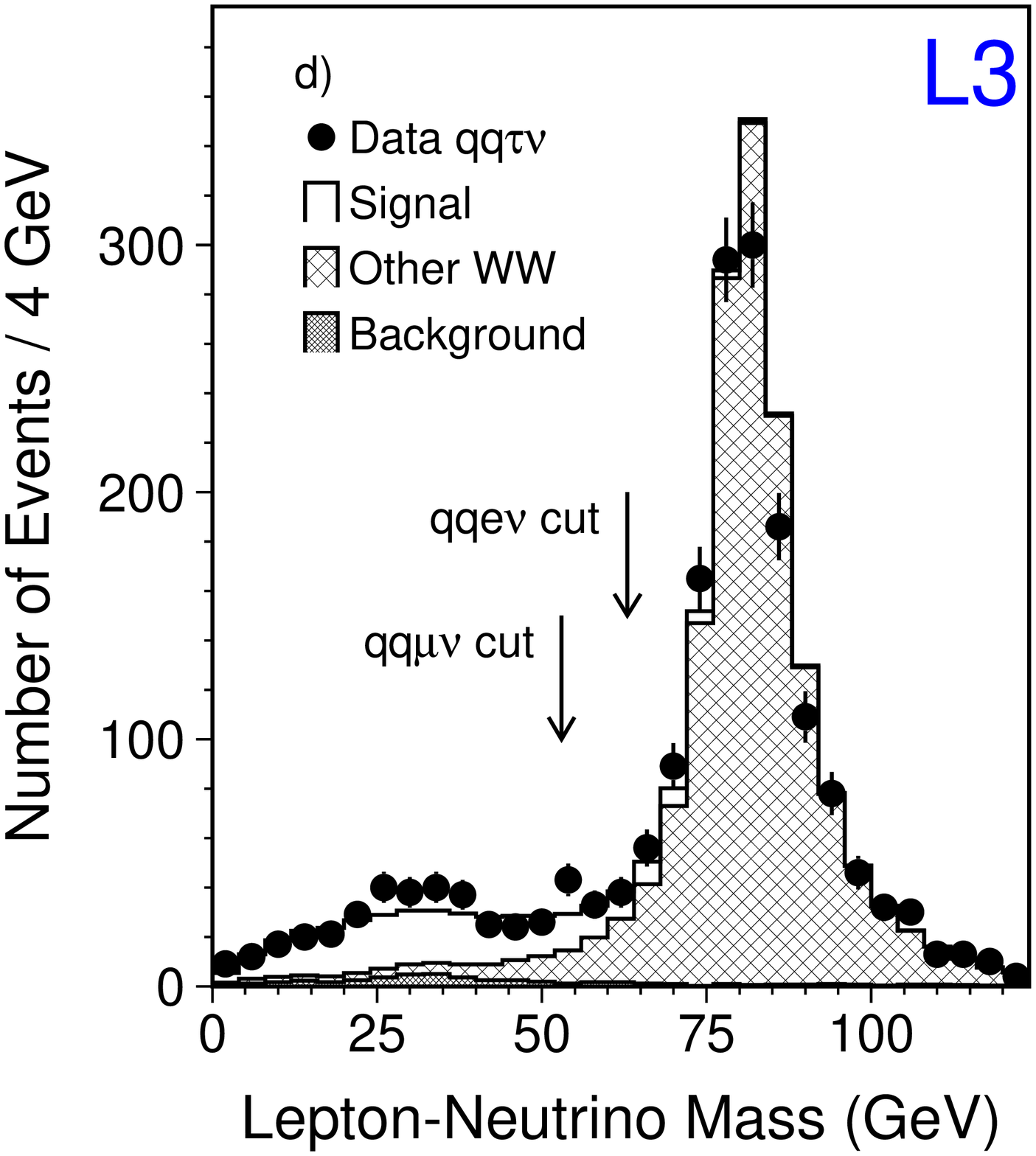,bb=10 30 560 650,width=0.5\linewidth}}
\end{tabular}
\caption[]{
  Distributions of variables used for the selection of $\QQMN$ and $\QQTN$
  events, comparing the signal and background Monte Carlo to the
  data collected at $\sqrt{s}=189-209\, \GeV$. 
 (a) The invariant mass of the jet-jet system in $\QQMN$ events.
 (b) The quantity $\psi_{\mu j}\times \sin\theta_\mathrm{miss}$ in $\QQMN$ events. 
 (c) The invariant mass of the jet-jet system in
 $\QQTN$ events.  
 (d) The invariant mass of the lepton-neutrino system for leptonically 
     decaying tau candidates  in $\QQTN$ events.
}
\label{fig:qqmnqqtn}
\end{center} 
\end{figure}

The purity of the selection is 98\% at $\sqrt{s}=206.5\, \GeV$. The
residual background, not originating from W-boson pair production,
is dominated by Z-boson pair-production events
(52\%) and $\EEQQG$ events (31\%).
The distributions of $M_\mathrm{jj}$ and of 
$\psi_{\mu j}\times \sin \theta_\mathrm{miss}$ are shown in Figure~\ref{fig:qqmnqqtn}.

\subsection{The \boldmath\protect$\QQTN$ Selection}
\label{sec:qqtn}

The event selection for the process $\EEQQTN$ is based on the
identification of an isolated low-momentum electron, muon, or narrow jet
in a hadronic environment with large missing energy.

Events are selected requiring $P_\mathrm{t}>10\,\GeV$,
$30~\GeV < M_\mathrm{jj} < 110~\GeV$ and the mass recoiling 
against the two-jet system  to be greater than $35~\GeV$.

Events are classified according to the presence of isolated electrons
or muons with an energy of more than $5~\GeV$. MIPs are not
considered as tau candidates.

For leptonically-decaying tau candidates,
cuts on $M_{\mathrm{e}\nu}$ and $M_{\mu\nu}$, complementary to those   
described in Sections~\ref{sec:qqen} and~\ref{sec:qqmn}, are applied. 
These cuts are chosen so as to minimise 
correlations in the measured W-boson branching fractions.

If no electrons or
muons are found, a search for a tau-jet is performed using a neural
network which exploits the distinctive characteristics of a hadronic
tau decay: low multiplicity, small jet opening angle, low jet mass and
high electromagnetic fraction of the jet energy. 
The jet with the highest
neural-network output is retained as the tau-lepton candidate.
For these events, additional requirements are applied 
in order to reduce the dominant background from $\EEQQG$ events. If
$P_\mathrm{t}<20~\GeV$,
the neural-network output of the tau-jet candidate is required 
to be near to that expected for a tau-jet.
At most three charged tracks are allowed to form the tau-jet
candidate. The polar angle of the missing momentum must satisfy 
$|\cos \theta_\mathrm{miss}| < 0.91$. The solid angle
subtended by the directions of the tau-jet candidate and the other two jets must be
less than 6~sr.

Among the events selected at $\sqrt{s}=206.5\, \GeV$, 62\% come from $\EEQQTN$ W-boson
pair-production processes and 21\% from other final states of the W-boson
pair production.
The background is dominated   
by $\EEQQG$ events (54\%) and $\EEQQEN$ final states not
originating from W-boson pair production (46\%).
The distributions of $M_{\ell\nu}$ and
$M_\mathrm{jj}$ are shown in Figure~\ref{fig:qqmnqqtn}.

\subsection{The \boldmath\protect$\QQQQ$ Selection}
\label{sec:qqqq}

The event selection for the process $\EEQQQQ$ requires hadronic events 
with little missing energy, 
high multiplicity and a four-jet topology. 

The Durham jet-resolution parameter $y_{34}$, for which the event topology changes
from three to four jets, is required to be greater than
$0.0015$. 
The events are clustered into four jets and a kinematic fit,
assuming four-momentum conservation, is used to improve energy and
angle resolutions. 

A neural network is 
trained to discriminate against the dominant $\EEQQG$ background.
Ten variables are used in the neural network: 
the spherocity~\cite{spherocity}, the lowest jet-multiplicity,  
$y_{34}$, the energies of the most and of the least
energetic jets, the difference between the energies
of the second and the third most energetic jets, the
broadenings~\cite{broadness} of the most and of the least energetic
jets, the probability of the kinematic fit  and the sum of the cosines of the six
angles between the four jets.
The dominant background 
is due to $\EEQQG$ events with four reconstructed jets, mainly
coming from $\EE \rightarrow \QQ \mathrm{gg}$ events.  We find that the four-jet rate in  $\EEQQG$ events is not well described
   by our MC simulations, and a comparison with data 
is used to determine this background.
Data and Monte Carlo distributions of the $y_{34}$ variable 
in hadronic Z decays collected at
$\sqrt{s}=91~\GeV$ are compared and their ratio 
is used to reweight the $\EEQQG$ Monte Carlo
events at higher energies throughout the rest of the analysis.
The resulting accepted number of
$\EEQQG$ events, for a neural network output greater than 0.6, is
increased by 12.7\%. 

\begin{figure}[p]
\begin{center}
{\epsfig{file=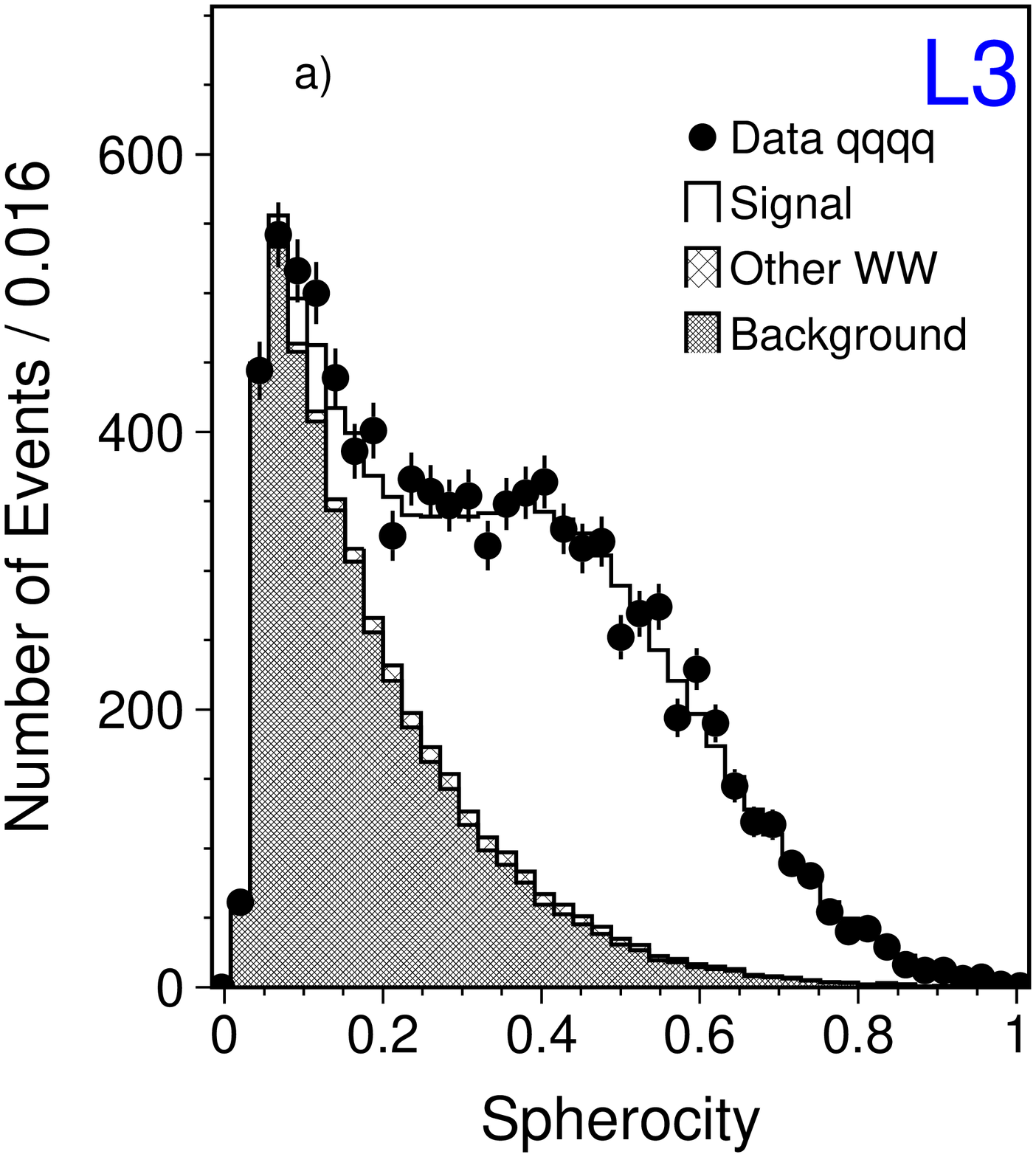,bb=10 30 560 650,width=0.49\linewidth}}
{\epsfig{file=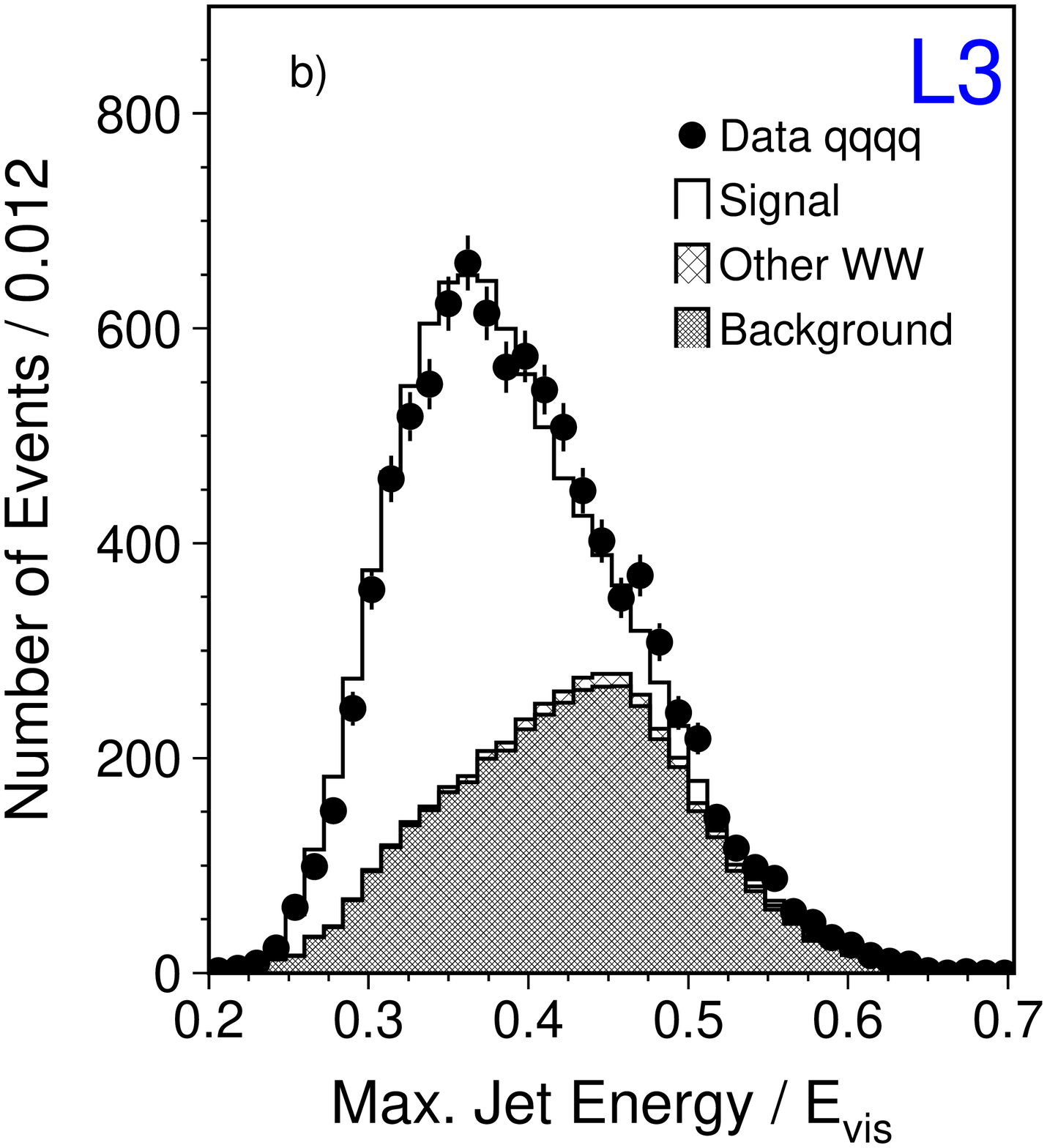,bb=10 30 560 650,width=0.49\linewidth}}\\
{\epsfig{file=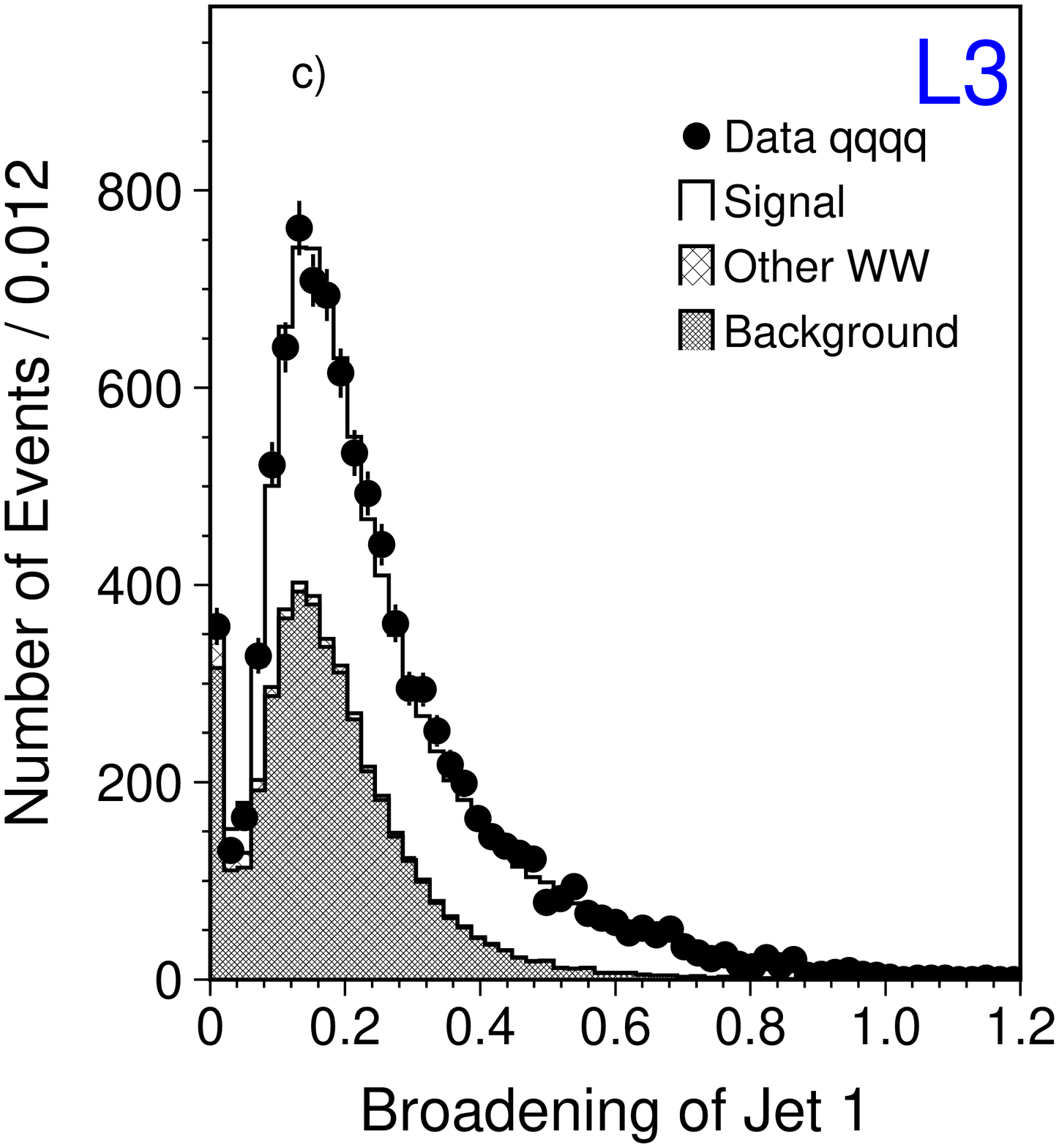,bb=10 30 560 650,width=0.49\linewidth}}
{\epsfig{file=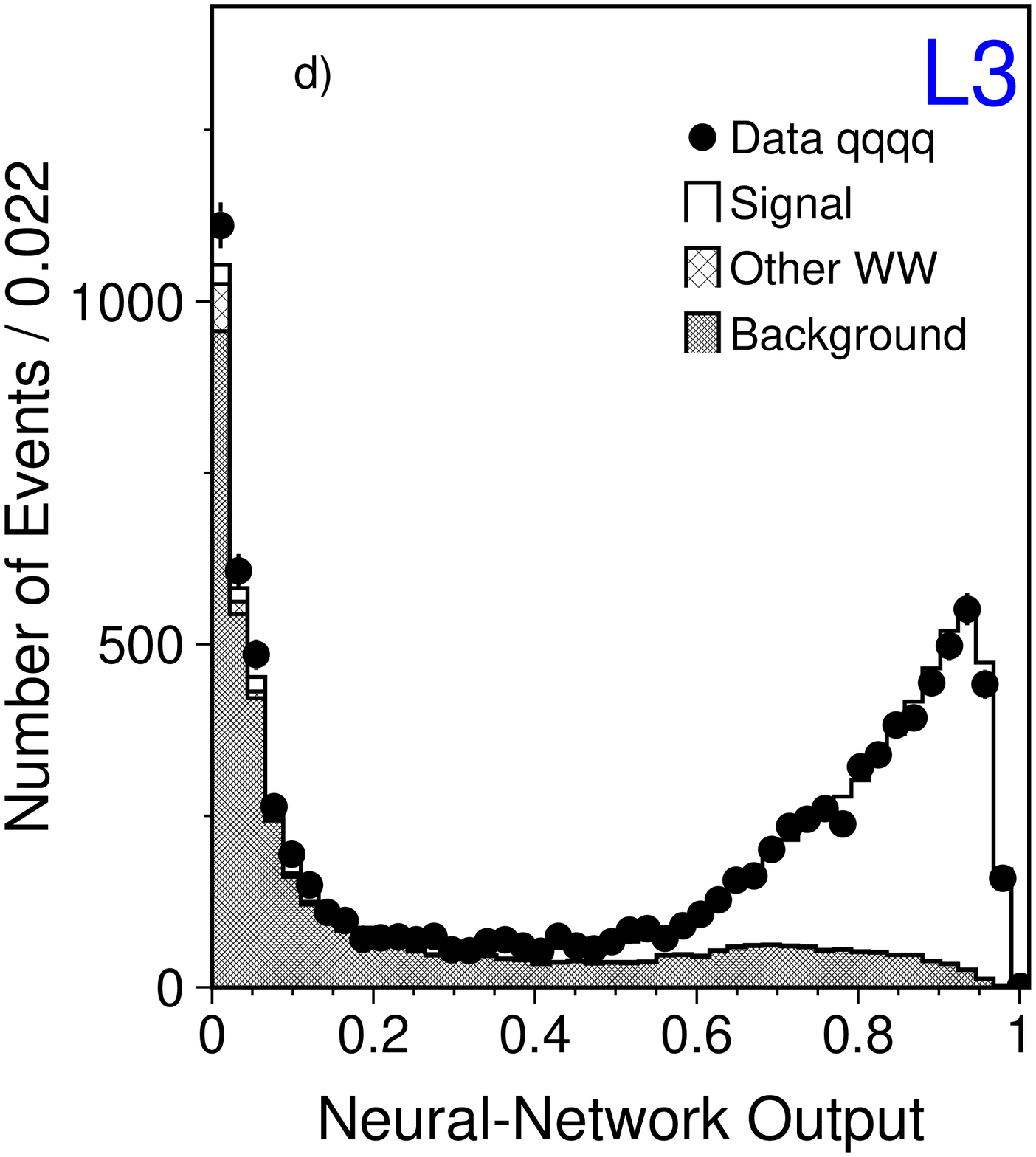,bb=10 30 560 650,width=0.49\linewidth}}
\caption[]{
  Distributions of some of the variables used for the neural network in the
  analysis of $\QQQQ$ events together with the neural-network output,
  comparing the signal and background Monte Carlo to the
  data collected at $\sqrt{s}=189-209\, \GeV$.
  (a) The spherocity.
  (b) The maximum jet energy scaled by the visible energy.
  (c) The broadening of the most energetic jet.
  (d) The neural-network output.
}
\label{fig:qqqq-nnin}
\end{center}
\end{figure}

Requiring the neural-network output to be greater than 0.6 yields a
sample purity of 80\% at $\sqrt{s}=206.5\, \GeV$ with a background dominated by  
the $\EEQQG$ (59\%) and Z-boson pair-production (41\%) processes. 
The distributions of some of the neural-network inputs and of
its output, peaking at one for the signal and at zero for the
background, are shown in Figure~\ref{fig:qqqq-nnin}.

\section{Fit Method}
\label{sec:fitmethod}
The CC03-level cross sections, $\sigma_j$, of the signal processes $j$ are
determined simultaneously in a single maximum-likelihood fit, taking 
cross-feed between different final states into account.  

For the purely leptonic final states, the fit procedure determines
six different cross sections corresponding to all possible
lepton-flavour combinations.  Since
the statistics for the $\LNLN$ final state is low, the sum of
these six cross sections is quoted in the following as the cross section for the
process $\EELNLN$.

The total likelihood is given by the product of Poissonian
probabilities, $P(N_i,\mu_i)$, to observe $N_i$ events in the $i$-th
final state, as listed in Table~\ref{tab:xsec1}.
The expected number of events for selection $i$, $\mu_i$, is calculated as:
\begin{eqnarray}
\mu_i & = &  
\left(\sum_{j=1}^{10}\epsilon_{ij}\sigma_j + 
\sum_{k=1}^{N_i^\mathrm{bg}} \epsilon_{ik}^\mathrm{bg} \sigma_k^\mathrm{bg}
\right) {\cal L}\,,
\label{eq:poisson-mu}
\end{eqnarray}
where $\epsilon_{ij}$ is the CC03-level efficiency of selection $i$ to accept
events from process $j$, $\sigma_k^\mathrm{bg}$ is the cross section
of the $k$-th background process, selected  with efficiency $\epsilon_{ik}^\mathrm{bg}$.
The $N_i^\mathrm{bg}$ background processes for selection $i$ also
include four-fermion final states not originating from W-boson pair production. 

For the
$\EEQQQQ$ process, the Poissonian probability is replaced by the
likelihood as a function of the signal cross section derived from a
fit to the neural-network output distribution. 
In this fit the $\EEQQG$ background contribution is fixed to the
value derived directly from data by performing a fit with both the signal and
background normalisations left free.
The results for the $\EEQQG$ background cross sections are shown in
Table~\ref{tab:xsecqq}. These values are in good agreement with
the Monte Carlo predictions. 
As a cross-check, the $\EEQQQQ$ cross section  is also determined by
repeating the full fit after applying a
cut on the output of the neural network at 0.6, which
minimises the expected statistical uncertainty. All values agree well
with those derived from the neural-network fit.  

\section{Systematic Uncertainties}
\label{sec:syst}

In addition to the uncertainty on the luminosity
measurement~\cite{l3-lumi96} and that due to  limited Monte Carlo statistics,
which affect all final states in common, the remaining
sources of systematic effects in the measurement of
W-boson pair-production cross sections are divided into two classes: 
uncertainties in the detector response and 
theoretical uncertainties. The latter come mainly from the knowledge
and modelling of the hadronisation processes.
A summary of the systematic uncertainties from all considered  sources is given in
Table~\ref{tab:syst} for $\sqrt{s}=206.5\, \GeV$. Values at different
$\sqrt{s}$ are only marginally different. Details about the assessment
of the systematic uncertainties are discussed below.

A possible source of systematic uncertainty arises from the accuracy
of the Monte Carlo modelling of the detector response. For the
semi-leptonic and fully-leptonic final states, this uncertainty is evaluated by varying
the positions of the selection cuts for each channel. The variation of
the cut positions is chosen so as to span several times the resolution
of the studied variable. Each variable is considered in turn and the
corresponding change in the measured cross sections are evaluated. For
variables which are correlated, for instance visible energies and
transverse momenta, the largest variation is retained. For the
selected variables, the expected statistical uncertainty on the
newly-selected data sample is subtracted from the observed variation
and the sum in quadrature of all results is retained as systematic
uncertainty.
Most of the systematic effects are related to the resolution of the
missing momentum.  In addition, the electron/photon discrimination
represents also a sizable source of systematic uncertainty for the
$\QQEN$ final state.

For the $\QQQQ$ selection, the systematic uncertainty on the
neural-network output is estimated by re-evaluating the input
variables of the neural network after smearing and scaling the
measurements of energy depositions and tracks in the simulation
according to the uncertainties on their resolutions.

The relative systematic uncertainty on the measured cross section,
assigned to detector response and modelling, varies from 1.0\% to
2.0\% depending on the final state.

As a cross-check, changes in efficiency due to variations of the
detector calibration within its uncertainty, are also studied. The
calibration is studied using samples of di-lepton and di-jet events,
collected during the calibration runs at $\sqrt{s}=91~\GeV$ and at
higher energies. The results of this study show a much smaller effect
than the cut-variation technique. The trigger inefficiency, as well as
its uncertainty, is found to be negligible in all channels.

Fragmentation and hadronisation uncertainties may affect both the
signal efficiency and the $\EEQQG$ background estimation. 
The modelling of the signal hadronisation is studied comparing the
selection efficiencies obtained with different
hadronisation models: PYTHIA, HERWIG~\cite{HERWIG} and
ARIADNE~\cite{ARIADNE}.  
The average difference with respect to PYTHIA gives  
a systematic uncertainty on the measured cross section of 
0.5\% to 1.2\%, dependent on the final state.

The effect of the hadronisation uncertainty in $\EEQQG$ background
events is also studied by comparing PYTHIA, HERWIG and ARIADNE. It is
found to be negligible for $\QQLN$ final states. In the $\QQQQ$ final
state, the hadronisation uncertainty affects mainly the four-jet rate
as described in Section~\ref{sec:qqqq}. Half of the effect due to the
$y_{34}$ reweighting is assigned as systematic uncertainty. It
corresponds to 0.9~\% of the measured $\EEQQQQ$ cross section.

Other sources of theoretical uncertainties in the $\QQQQ$ channel
arise from correlations among final-state hadrons such as
Bose-Einstein correlations and colour reconnection.
The modelling of Bose-Einstein correlations between hadrons from W-boson
decays may affect the selection efficiencies.  
In previous studies~\cite{l3-257} we have measured the
strength of Bose-Einstein correlations between hadrons originating
from the same W boson in semi-leptonic W decays. 
Its value is significantly different from zero and in good agreement with
that for light-quark Z decays and also with 
that of the LUBOEI BE$_{32}$ model~\cite{LUBOEI} used in our Monte Carlo
simulations.
The systematic uncertainty derived from the uncertainty of this
strength is found to be negligible.
Bose-Einstein correlations between particles originating from different W bosons are
strongly disfavoured in $\EEQQQQ$ events~\cite{l3-257}. 
Their measured strength is restricted to
at most a quarter   of the strength expected in the BE$_{32}$
model with full correlations. Allowing  correlations with such a
strength  yields negligible changes in the measured cross sections.  

Extreme models of colour reconnection between the hadronic
systems in $\QQQQ$ events are disfavoured by
data~\cite{l3-266,l3-272}. The influence of colour reconnection is estimated using
the models implemented in HERWIG~\cite{HERWIG_cr},
ARIADNE~\cite{ARIADNE_cr} (model 1 and model 2)
and PYTHIA (model SK I with reconnection
parameter $k=0.6$~\cite{SKMODELS}). The ARIADNE-2 model is compared to a modified version 
of the ARIADNE-1 model, so that in both models the shower cascade is 
performed in two phases with an identical energy cut-off parameter.
The average difference of 0.19\% with respect to PYTHIA  is assigned
as systematic uncertainty on the measured $\EEQQQQ$ cross section.

The theoretical uncertainties on the cross sections of the background processes, namely 
hadron production in two-photon collisions (50\%), neutral-current four-fermion
processes (5\%) and fermion-pair production (1\%) lead to systematic
uncertainties of 0.1\% to 0.4\%. In the determination of the $\EEQQQQ$
cross sections, the $\EEQQG$ background levels are directly measured from
data and the corresponding uncertainties, as reported in
Table~\ref{tab:xsecqq}, are propagated to the final results.

The dependence of the selection efficiencies on the mass and width of
the W boson, $\MW$ and $\GW$, is studied using Monte Carlo samples simulated with
different $\MW$ and $\GW$ values. The propagation of the world-average
uncertainties on these two parameters, 40~$\MeV$ on $\MW$ and
60~$\MeV$ on $\GW$~\cite{PDG02}, is taken as systematic uncertainty.
It corresponds to a less than 0.3\% effect.

The systematic uncertainty on initial-state radiation (ISR), due to its
approximate leading-log ${\cal O}(\alpha^3)$ treatment in KandY, is
investigated by re-evaluating the signal efficiencies for Monte Carlo
events reweighted by
$|{\cal M}[{\cal O}(\alpha^2)]|^2/|{\cal M}[{\cal O}(\alpha^3)]|^2$. 
The effect is found
to be negligible. As a cross-check, the Monte Carlo events are also
reweighted by 10\% in the presence of ISR photons with energies or
transverse momenta exceeding 100~$\MeV$. In both cases, the effect
is negligible.

Final-state radiation (FSR) is 
implemented in  KandY using the PHOTOS package~\cite{PHOTOS} based
on the leading-log approximation. 
The PHOTOS package is inaccurate in the hard non-collinear region.
The related systematic
uncertainty is estimated by determining  the selection efficiencies 
using Monte Carlo events whose weights are reduced by 50\% in the presence of FSR photons 
with energy greater than $30~\GeV$. An effect between 0.1\% and 0.2\%
is observed and retained as systematic
uncertainty.

Uncertainties due to the implementation of virtual $\cal{O}(\alpha)$
corrections in the KandY program are tested
comparing signal efficiencies to those obtained with the
RacoonWW program. No sizable effect is observed.

Correlations among all sources of systematic uncertainties are taken
into account in the following results.

\section{Results}
\label{sec:results}

\subsection{Single-Channel Cross Sections}
Fits are performed to derive ten cross sections, one for each final state.
No assumption is made concerning the W-boson branching fractions.
The results, summing up all fully leptonic final
states and including statistical and systematic
uncertainties, are listed in Table~\ref{tab:xsec1}.  The Standard Model agrees
well with these results.  Since the efficiency matrix of
Table~\ref{tab:xmat-1} contains non-zero off-diagonal elements, the
measured cross sections are correlated.  The largest correlations,  $-10.3\%$
and  $-17.6\%$, are between the  $\EEQQTN$ and
$\EEQQEN$ and between the $\EEQQTN$ and $\EEQQMN$ cross
sections, respectively.  All other correlations are
less than 1\%.

\subsection{Total Cross Section and Branching Fractions}

For the determination of the CC03 cross section of
W-boson pair production, $\SWW$, the signal cross sections $\sigma_j$ are replaced by
the product $r_j\SWW$. The ratios $r_j$ are given in terms of the
W-boson decay
branching fractions, $\mathrm{Br}(\WQQ)$ and $\mathrm{Br}(\WLN)$, as follows:
$r_{\QQQQ}=[\mathrm{Br}(\WQQ)]^2$, $r_{\QQLN}=2\mathrm{Br}(\WQQ)\mathrm{Br}(\WLN)$, and $r_{\LNLN} =
[\mathrm{Br}(\WLN)]^2$ for same-flavour leptons or
$2\mathrm{Br}(\WLN)\mathrm{Br}(\mathrm{W}\rightarrow\ell'\nu)$
otherwise.

\begin{figure}[p]
\begin{center}
\begin{tabular}{c}
{\epsfig{file=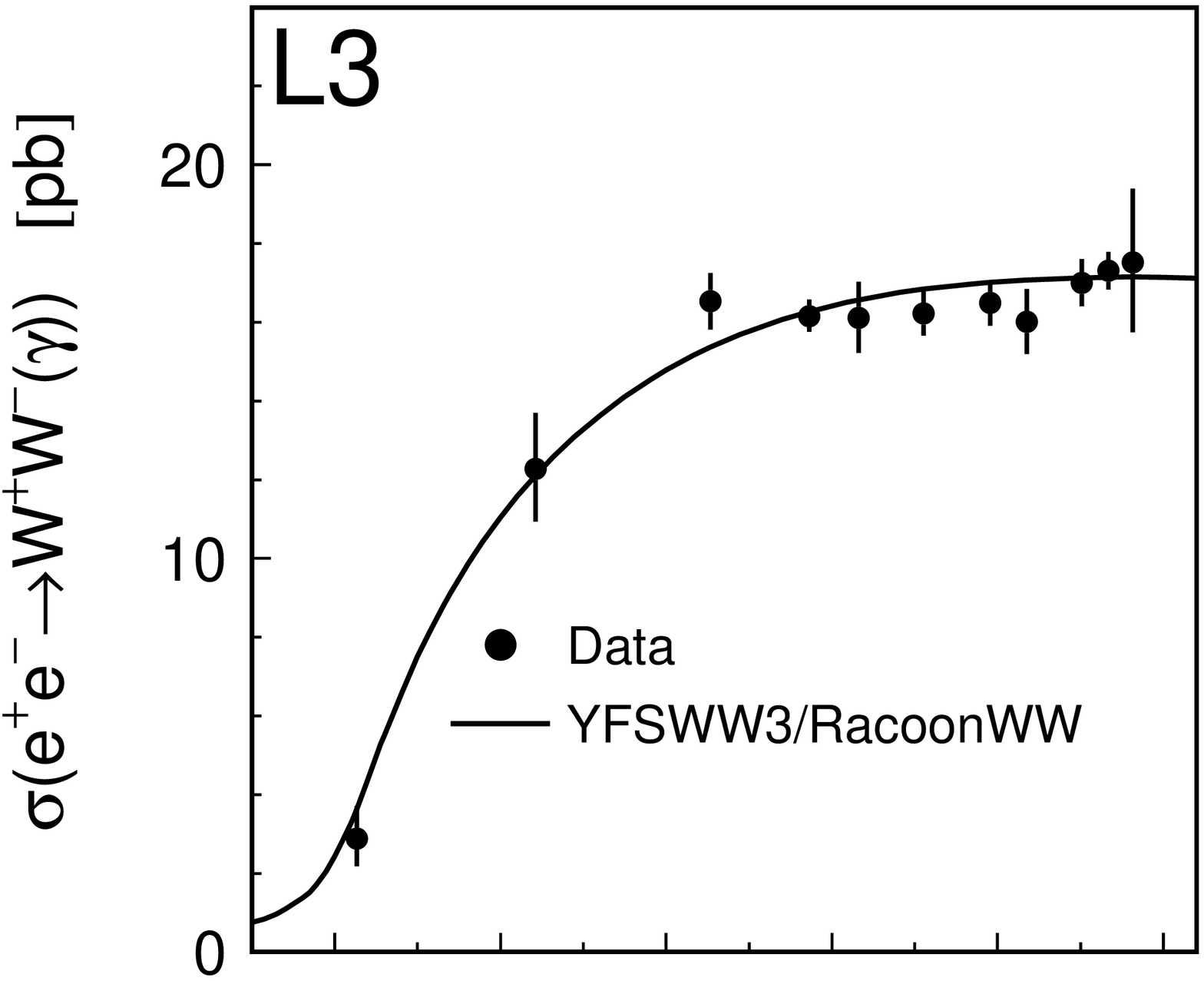,bb=0 134 540 560,width=0.9\linewidth}}\\
{\epsfig{file=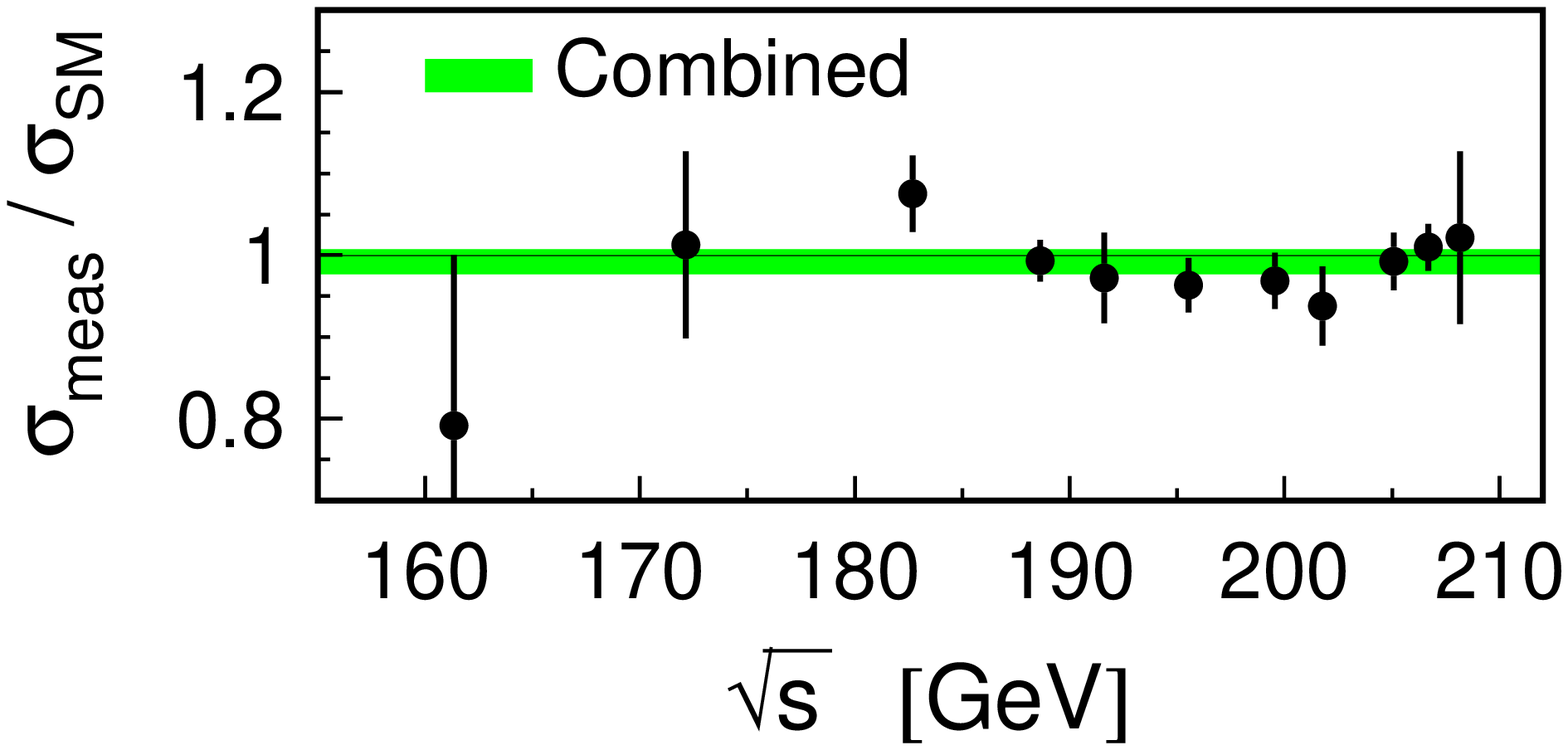,bb=0 40 540 299,width=0.9\linewidth}}
\end{tabular}
\caption[]{The cross section of the process
  $\EE \rightarrow \mathrm{W^+W^-}$ as a function of $\sqrt{s}$.
  The published measurements of $\SWW$ at $\sqrt{s}$ of $161~\GeV$,
  $172~\GeV$ and $183~\GeV$,  the updated
  measurement 
  at $\sqrt{s}=189~\GeV$ and the new 
  measurements at $\sqrt{s}=192-209~\GeV$ are shown 
  as dots with error bars, combining
  statistical and systematic uncertainties in quadrature.  The 
  solid curve shows the Standard Model expectation as calculated with
  YFSWW3 in the whole energy range and
  RacoonWW for $\sqrt{s}\ge 170~\GeV$. Its uncertainty of 0.5\% is
  invisible on this scale.  
  The lower plot shows the ratios of the measured cross sections with
  respect to the Standard Model expectations as calculated with YFSWW3.
  The band represent their combined value with its total uncertainty:
  $R = 0.992 \pm 0.015$.}
\label{fig:xsec}
\end{center}
\end{figure}

Results for the cross sections of the reactions
$\EELNLN$, $\EEQQLN$ and $\EEQQQQ$, assuming charged-lepton universality,
are obtained as shown in Table~\ref{tab:sigma3}.
The total cross sections, $\SWW$, are then derived assuming the Standard
Model W-boson decay branching fractions~\cite{LEP2YRWW} and are also
reported in Table~\ref{tab:sigma3} together with the Standard Model expectations.
Our previous measurements at $\sqrt{s}$ of $161~\GeV$~\cite{l3-111},
$172~\GeV$~\cite{l3-120}, $183~\GeV$~\cite{l3-155} and these results  
are compared in Figure~\ref{fig:xsec} 
to the Standard Model expectation as calculated with the Monte Carlo programs
YFSWW3 and RacoonWW.
The two predictions agree with our data and are consistent within a theoretical uncertainty
of 0.5\%~\cite{MCWKSHOP} for $\sqrt{s}\ge 170~\GeV$.

The ratios of the measured cross sections to the Standard
Model predictions of the YFSWW3 program are also shown in Figure~\ref{fig:xsec}.
Their combined value, $R$, is:
\begin{eqnarray}
R = 0.992 \pm 0.011 \pm 0.009  \pm 0.005,
\end{eqnarray}
where the first uncertainty is statistical, the second systematic and
the third theoretical.

For the determination of W-boson decay branching fractions, the data
collected at lower centre-of-mass energies are also included.
The sum of the hadronic
and the three leptonic branching fractions is constrained to unity.
The branching fractions are
first determined without the assumption of charged-lepton
universality, with the results listed in Table~\ref{tab:brafra}. 
The hypothesis of charged-lepton
universality is tested and  the probability of getting a $\chi^2$ greater than that
observed is 0.8\% differing by 2.6
standard deviations from this hypothesis.
Assuming charged-lepton universality,
the hadronic W-boson decay branching fraction is:
\begin{eqnarray}
\mathrm{Br}(\WQQ)        & = & 67.50\pm0.42\pm0.30~\%         \,,
\end{eqnarray}
where the first uncertainty is statistical and the second systematic.  
The W-boson decay branching fractions depend on the six elements
$V_{ij}$ of the Cabibbo-Kobayashi-Maskawa  matrix
$V$~\cite{VCKM} not involving the top
quark~\cite{LEP2YRWW}:
\begin{eqnarray*}
1/\mathrm{Br}(\WLN) & = & 3+3[1+\alpha_{\mathrm{s}}(\MW)/\pi]
\sum_{i=\mathrm{u,c};\,j=\mathrm{d,s,b}} \left|V_{ij}\right|^2,
\end{eqnarray*}
where $\alpha_{\mathrm{s}}$ is the strong coupling constant.  
Using $\alpha_{\mathrm{s}}=0.119\pm0.002$~\cite{PDG02}, 
our measurements correspond to:
\begin{eqnarray}
 \sum_{i=\mathrm{u,c};\,j=\mathrm{d,s,b}} \left|V_{ij}\right|^2 & = &  2.002 \pm 0.038 \pm 0.027 \,,
\end{eqnarray}
where the first uncertainty is statistical and the second systematic.

Using the current world-average values and uncertainties of the other matrix
elements, not imposing the unitarity of
the $V$ matrix, $|V_{\mathrm{cs}}|$ is derived as:
\begin{eqnarray}
  |V_{\mathrm{cs}}| & = & 0.977 \pm 0.020 \pm 0.014 \,,
\end{eqnarray}
where the first uncertainty is statistical and the second systematic.
The systematic uncertainty includes the uncertainties on
$\alpha_{\mathrm{s}}$ and on the other matrix elements~\cite{PDG02}.

\subsection{Differential Cross Section}
The combined differential cross section for the $\EEQQEN$ and $\EEQQMN$
channels, as a function of
$\ctw$, where $\theta_{\mathrm{W^-}}$ is the $\mathrm{W^-}$ production
angle with respect to the direction of the incoming electrons, is
measured for different $\sqrt{s}$ from $183\,\GeV$ to $209\,\GeV$.
These two channels are used because  the lepton charge tags
the W-boson charge with high purity. 

Four energy bins are considered:
\begin{center}
\begin{tabular}{c}
{ \bf $180.0 - 184.0~\GeV$}, { $184.0 - 194.0~\GeV$} \\
~~{ \bf $194.0 - 204.0~\GeV$}, { $204.0 - 209.0~\GeV$} .\\
\end{tabular}
\end{center}

\noindent
These are chosen so as to minimise the difference between the
average slope of the differential cross section in each bin and the slope corresponding to the
luminosity-weighted average centre-of-mass energies: 
$\langle \sqrt{s} \rangle = 182.7,
189.0, 198.3$ and $205.9$ $\GeV$, respectively.
In each energy range, ten $\ctw$ bins are studied. 
The variable $\ctw$ is reconstructed from the measurements of the jet
and lepton angles and energies~\cite{l3-284}. Monte Carlo events are
then used to extract the differential cross section. Ambiguities might
arise in the presence of additional photons in the generated events,
and the $\ctw$ angle is then defined following the
$\gamma$-recombination scheme~\cite{MCWKSHOP}:
\begin{itemize}
\item all photons inside a cone of 5 degrees half-opening angle with respect to
  the beam direction are treated as invisible;
\item the combined mass of each photon with electrons, muons and
quarks is calculated. If the smallest combined mass is less than
  $5~\GeV$ or the energy of the photon is less than $1~\GeV$, the
  momentum of the photon is added to that of the fermion and the
  photon is discarded.
\end{itemize}
The measured cross sections are corrected to CC03-level 
with the additional restriction of
$20^\circ<\theta_{\ell^{\pm}}<160^\circ$, 
where $\theta_{\ell^{\pm}}$ is the angle between the charged
lepton and the beam direction. 

\begin{figure}[p]
\begin{center}
{\epsfig{file=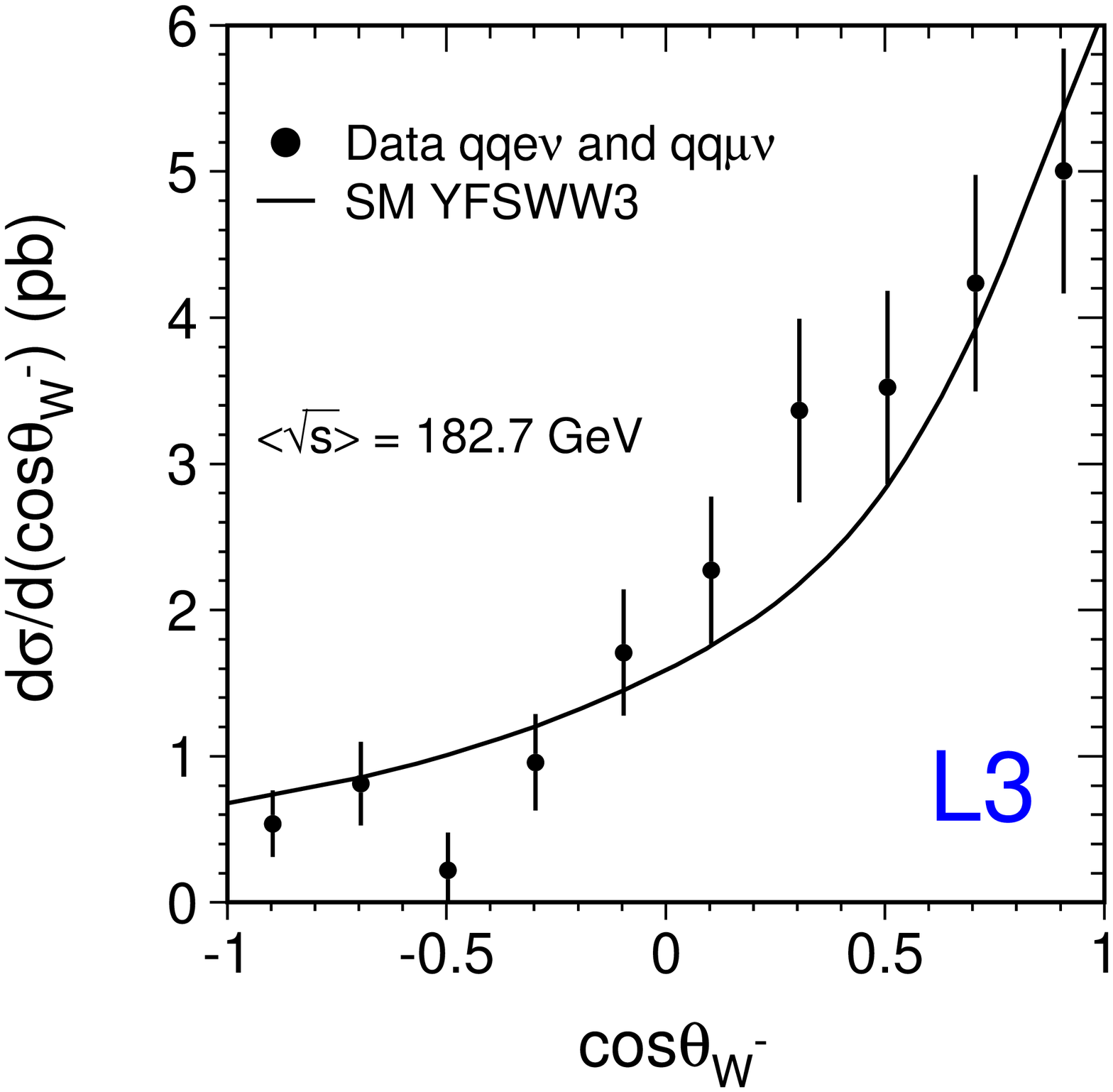,width=0.49\linewidth}}
{\epsfig{file=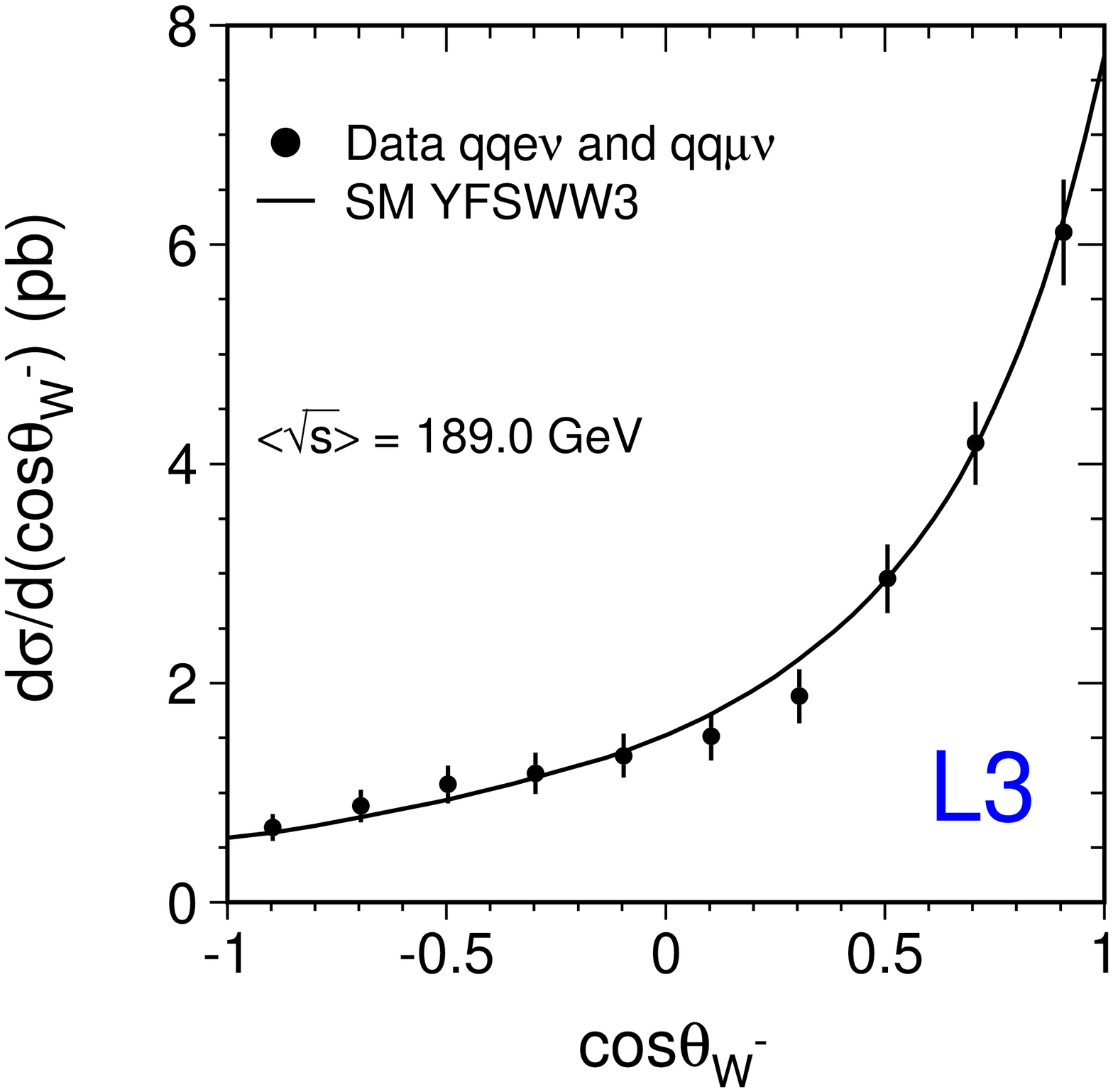,width=0.49\linewidth}}\\
{\epsfig{file=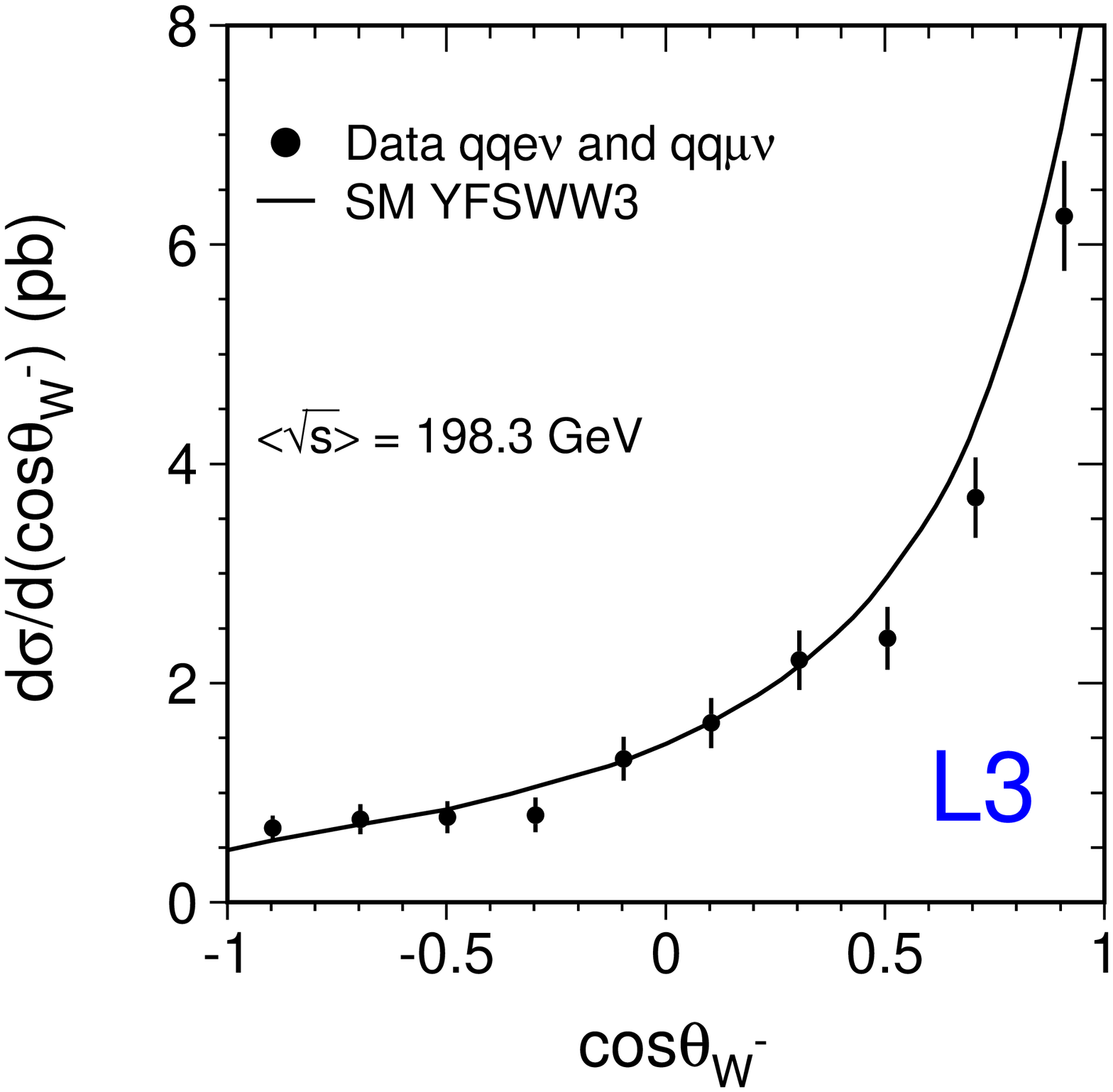,width=0.49\linewidth}}
{\epsfig{file=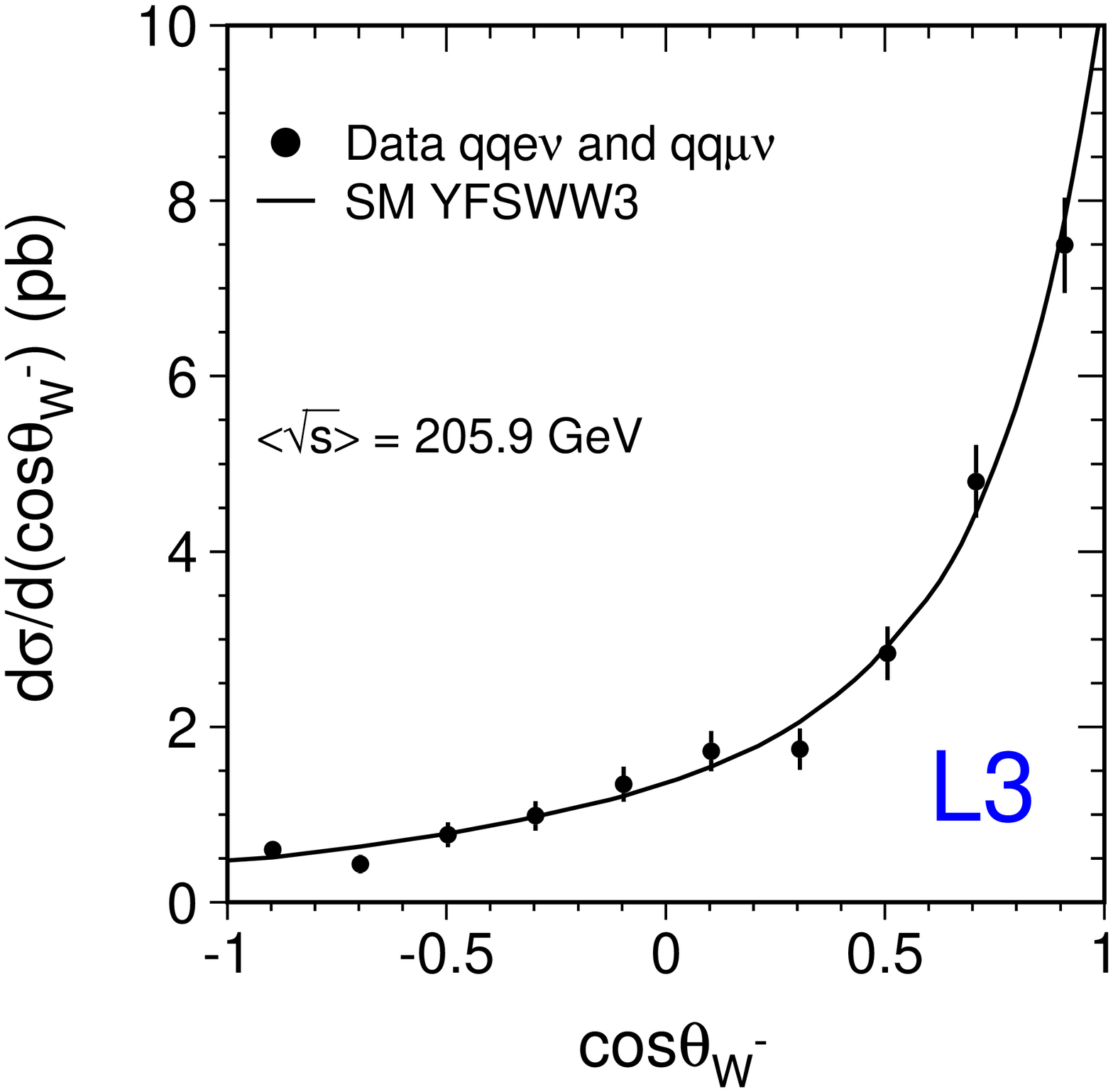,width=0.49\linewidth}}
\caption[]{Measured differential cross sections as a function of
$\ctw$ for the $\EEQQEN$ and $\EEQQMN$ processes. The cross
sections of the two channels are summed. Experimental
data are represented by dots with error bars which include statistical
and systematic uncertainties added in quadrature. Monte Carlo
expectations are shown as  solid lines.  
}
\label{fig:ds}
\end{center}
\end{figure}

The observed $\ctw$ distributions are corrected to
generator level, after background subtraction, by 
using bin-by-bin correction factors and
the cross sections in each $\ctw$ bin are determined as listed in 
Table~\ref{tab:diffxs} and plotted in Figure~\ref{fig:ds}.

As a cross-check, a full matrix unfolding from reconstruction to generator
level is also used. Since the migration matrix is almost diagonal, with
bin-to-bin migration effects at the level of 20\% at most, the
results are in perfect agreement with the simple bin-by-bin
correction method.

The potential bias of implicitly
assuming the Standard Model $\ctw$ distribution in the correction
factors, is studied using simulated samples with modified $\ctw$
behaviour and found to be negligible.
Another bias could arise directly from the W-boson pair-production Monte Carlo
generator used to estimate the correction factors.
No difference between KORALW and YFSWW3 programs is observed, hence no
additional systematic uncertainty is assigned.

Charge-confusion effects, which affect the reconstruction of the
W-boson direction, are taken into account. The residual uncertainty,
obtained by comparing data and Monte Carlo expectations on Z-peak
samples~\cite{l3-284}, is retained as a systematic uncertainty in
addition to those affecting the
total W-boson pair-production  cross section.

The systematic uncertainty 
is taken to be fully correlated between $\ctw$ bins and
energy points. 

%
\section{Conclusions}

In a data sample corresponding to an integrated luminosity of
629.4~pb$^{-1}$, collected at centre-of-mass energies ranging from
189~$\GeV$ to 209~$\GeV$, W-boson pair-production cross sections are
measured by selecting four-fermion events 
and found to be in agreement with Standard Model expectations.

The branching fractions for leptonic W decays are measured for each
lepton generation. 
Assuming charged-lepton universality, the branching fraction for hadronic W decays is
measured to be: $67.50 \pm 0.42~(\mathrm{stat.}) \pm
0.30~(\mathrm{syst.})~\%$.  
Combining all $\sqrt{s}$ points,
the ratio $R$ of the measured total W-boson pair-production cross section with respect to the 
theoretical  prediction is
found to be $ R = 0.992 \pm0.011\mathrm{(stat.)} \pm 0.009\mathrm{(syst.)} \pm 0.005\mathrm{(theo.)}$. 

Differential cross sections as a  function of the $\mathrm{W^-}$
production angle are
also measured and found to be in good agreement with Standard Model predictions.

\clearpage
\vfill
%
%
\bibliographystyle{/l3/paper/biblio/l3stylem}
\bibliography{l3aty2k}

%
%
\newpage
\typeout{   }     
\typeout{Using author list for W Cross section paper #291 ! ! !   }
\typeout{$Modified: Jul 15 2001 by smele $}
\typeout{!!!!  This should only be used with document option a4p!!!!}
\typeout{   }
%
%
%
%
%
%

\newcount\tutecount  \tutecount=0
\def\tutenum#1{\global\advance\tutecount by 1 \xdef#1{\the\tutecount}}
\def\tute#1{$^{#1}$}
\tutenum\aachen            
\tutenum\nikhef            
\tutenum\mich              
\tutenum\lapp              
\tutenum\basel             
\tutenum\lsu               
\tutenum\beijing           
\tutenum\bologna           
\tutenum\tata              
\tutenum\ne                
\tutenum\bucharest         
\tutenum\budapest          
\tutenum\mit               
\tutenum\panjab            
\tutenum\debrecen          
\tutenum\dublin            
\tutenum\florence          
\tutenum\cern              
\tutenum\wl                
\tutenum\geneva            
\tutenum\hamburg           
\tutenum\hefei             
\tutenum\lausanne          
\tutenum\lyon              
\tutenum\madrid            
\tutenum\florida           
\tutenum\milan             
\tutenum\moscow            
\tutenum\naples            
\tutenum\cyprus            
\tutenum\nymegen           
\tutenum\caltech           
\tutenum\perugia           
\tutenum\peters            
\tutenum\cmu               
\tutenum\potenza           
\tutenum\prince            
\tutenum\riverside         
\tutenum\rome              
\tutenum\salerno           
\tutenum\ucsd              
\tutenum\sofia             
\tutenum\korea             
\tutenum\taiwan            
\tutenum\tsinghua          
\tutenum\purdue            
\tutenum\psinst            
\tutenum\zeuthen           
\tutenum\eth               

{
\parskip=0pt
\noindent
{\bf The L3 Collaboration:}
\ifx\selectfont\undefined
 \baselineskip=10.8pt
 \baselineskip\baselinestretch\baselineskip
 \normalbaselineskip\baselineskip
 \ixpt
\else
 \fontsize{9}{10.8pt}\selectfont
\fi
\medskip
\tolerance=10000
\hbadness=5000
\raggedright
\hsize=162truemm\hoffset=0mm
\def\r{\rlap,}
\noindent

P.Achard\r\tute\geneva\ 
O.Adriani\r\tute{\florence}\ 
M.Aguilar-Benitez\r\tute\madrid\ 
J.Alcaraz\r\tute{\madrid}\ 
G.Alemanni\r\tute\lausanne\
J.Allaby\r\tute\cern\
A.Aloisio\r\tute\naples\ 
M.G.Alviggi\r\tute\naples\
H.Anderhub\r\tute\eth\ 
V.P.Andreev\r\tute{\lsu,\peters}\
% ...
\newpage
\begin{list}{A}{\itemsep=0pt plus 0pt minus 0pt\parsep=0pt plus 0pt minus 0pt
                \topsep=0pt plus 0pt minus 0pt}
\item[\aachen]
 III. Physikalisches Institut, RWTH, D-52056 Aachen, Germany$^{\S}$
\item[\nikhef] National Institute for High Energy Physics, NIKHEF, 
     and University of Amsterdam, NL-1009 DB Amsterdam, The Netherlands
\item[\mich] University of Michigan, Ann Arbor, MI 48109, USA
\item[\lapp] Laboratoire d'Annecy-le-Vieux de Physique des Particules, 
     LAPP,IN2P3-CNRS, BP 110, F-74941 Annecy-le-Vieux CEDEX, France
\item[\basel] Institute of Physics, University of Basel, CH-4056 Basel,
     Switzerland
\item[\lsu] Louisiana State University, Baton Rouge, LA 70803, USA
\item[\beijing] Institute of High Energy Physics, IHEP, 
  100039 Beijing, China$^{\triangle}$ 
\item[\bologna] University of Bologna and INFN-Sezione di Bologna, 
     I-40126 Bologna, Italy
\item[\tata] Tata Institute of Fundamental Research, Mumbai (Bombay) 400 005, India
\item[\ne] Northeastern University, Boston, MA 02115, USA
\item[\bucharest] Institute of Atomic Physics and University of Bucharest,
     R-76900 Bucharest, Romania
\item[\budapest] Central Research Institute for Physics of the 
     Hungarian Academy of Sciences, H-1525 Budapest 114, Hungary$^{\ddag}$
\item[\mit] Massachusetts Institute of Technology, Cambridge, MA 02139, USA
\item[\panjab] Panjab University, Chandigarh 160 014, India
\item[\debrecen] KLTE-ATOMKI, H-4010 Debrecen, Hungary$^\P$
\item[\dublin] Department of Experimental Physics,
  University College Dublin, Belfield, Dublin 4, Ireland
\item[\florence] INFN Sezione di Firenze and University of Florence, 
     I-50125 Florence, Italy
\item[\cern] European Laboratory for Particle Physics, CERN, 
     CH-1211 Geneva 23, Switzerland
\item[\wl] World Laboratory, FBLJA  Project, CH-1211 Geneva 23, Switzerland
\item[\geneva] University of Geneva, CH-1211 Geneva 4, Switzerland
\item[\hamburg] University of Hamburg, D-22761 Hamburg, Germany
\item[\hefei] Chinese University of Science and Technology, USTC,
      Hefei, Anhui 230 029, China$^{\triangle}$
\item[\lausanne] University of Lausanne, CH-1015 Lausanne, Switzerland
\item[\lyon] Institut de Physique Nucl\'eaire de Lyon, 
     IN2P3-CNRS,Universit\'e Claude Bernard, 
     F-69622 Villeurbanne, France
\item[\madrid] Centro de Investigaciones Energ{\'e}ticas, 
     Medioambientales y Tecnol\'ogicas, CIEMAT, E-28040 Madrid,
     Spain${\flat}$ 
\item[\florida] Florida Institute of Technology, Melbourne, FL 32901, USA
\item[\milan] INFN-Sezione di Milano, I-20133 Milan, Italy
\item[\moscow] Institute of Theoretical and Experimental Physics, ITEP, 
     Moscow, Russia
\item[\naples] INFN-Sezione di Napoli and University of Naples, 
     I-80125 Naples, Italy
\item[\cyprus] Department of Physics, University of Cyprus,
     Nicosia, Cyprus
\item[\nymegen] University of Nijmegen and NIKHEF, 
     NL-6525 ED Nijmegen, The Netherlands
\item[\caltech] California Institute of Technology, Pasadena, CA 91125, USA
\item[\perugia] INFN-Sezione di Perugia and Universit\`a Degli 
     Studi di Perugia, I-06100 Perugia, Italy   
\item[\peters] Nuclear Physics Institute, St. Petersburg, Russia
\item[\cmu] Carnegie Mellon University, Pittsburgh, PA 15213, USA
\item[\potenza] INFN-Sezione di Napoli and University of Potenza, 
     I-85100 Potenza, Italy
\item[\prince] Princeton University, Princeton, NJ 08544, USA
\item[\riverside] University of Californa, Riverside, CA 92521, USA
\item[\rome] INFN-Sezione di Roma and University of Rome, ``La Sapienza",
     I-00185 Rome, Italy
\item[\salerno] University and INFN, Salerno, I-84100 Salerno, Italy
\item[\ucsd] University of California, San Diego, CA 92093, USA
\item[\sofia] Bulgarian Academy of Sciences, Central Lab.~of 
     Mechatronics and Instrumentation, BU-1113 Sofia, Bulgaria
\item[\korea]  The Center for High Energy Physics, 
     Kyungpook National University, 702-701 Taegu, Republic of Korea
\item[\taiwan] National Central University, Chung-Li, Taiwan, China
\item[\tsinghua] Department of Physics, National Tsing Hua University,
      Taiwan, China
\item[\purdue] Purdue University, West Lafayette, IN 47907, USA
\item[\psinst] Paul Scherrer Institut, PSI, CH-5232 Villigen, Switzerland
\item[\zeuthen] DESY, D-15738 Zeuthen, Germany
\item[\eth] Eidgen\"ossische Technische Hochschule, ETH Z\"urich,
     CH-8093 Z\"urich, Switzerland
\item[\S]  Supported by the German Bundesministerium 
        f\"ur Bildung, Wissenschaft, Forschung und Technologie.
\item[\ddag] Supported by the Hungarian OTKA fund under contract
numbers T019181, F023259 and T037350.
\item[\P] Also supported by the Hungarian OTKA fund under contract
  number T026178.
\item[$\flat$] Supported also by the Comisi\'on Interministerial de Ciencia y 
        Tecnolog{\'\i}a.
\item[$\sharp$] Also supported by CONICET and Universidad Nacional de La Plata,
        CC 67, 1900 La Plata, Argentina.
\item[$\triangle$] Supported by the National Natural Science
  Foundation of China.
\end{list}
}
\vfill


\newpage

\begin{table}[p]
\begin{center}
\renewcommand{\arraystretch}{1.2}
\begin{tabular}{|c|cccccccc|}
\hline
$\langle \sqrt{s} \rangle \, [\GeV]$ & 
188.6 & 191.6 & 195.5 & 199.6 & 201.8 & 204.8 & 206.5 & 208.0 \\
\hline
${\cal L} \, [\pb] $ &
176.8 & 29.8 & 84.1 & 83.3 & 37.1 & 79.0 & 130.5 & 8.6 \\
\hline 
\end{tabular}
\caption[]{
Average centre-of-mass energies and integrated
luminosities.}
\label{tab:stat}
\end{center}
\end{table}

\begin{table}[p]
\begin{center}
\renewcommand{\arraystretch}{1.2}
\begin{tabular}{| c || c c c c c c | c c c | c|}
\hline
\multicolumn{1}{|c||}{Selection} & 
\multicolumn{10}{|c|}{Efficiencies [\%] for $\mathrm{e^+e^-} \rightarrow$}  \\     
          &  $\ENEN$ &  $\ENMN$ &  $\ENTN$ &  $\MNMN$ &  $\MNTN$ & 
             $\TNTN$ &  $\QQEN$ &  $\QQMN$ &  $\QQTN$ &  $\QQQQ$ \\
\hline
\hline

$\ENEN$& $54.7$ &$\pz0.8$& $11.4$ &        &$\pz0.1$&$\pz1.5$&        &        &        &         \\
$\ENMN$&        & $47.6$ &$\pz8.4$&$\pz1.4$& $10.1$ &$\pz2.2$&        &        &        &         \\
$\ENTN$&$\pz6.0$&$\pz1.7$& $27.8$ &        &$\pz0.4$&$\pz7.5$&        &        &        &         \\
$\MNMN$&        &        &        & $41.0$ &$\pz6.9$&$\pz0.9$&        &        &        &         \\
$\MNTN$&        &$\pz2.6$&$\pz0.3$&$\pz3.0$& $23.1$ &$\pz4.8$&        &        &        &         \\
$\TNTN$&$\pz0.2$&$\pz0.1$&$\pz2.1$&        &$\pz1.3$& $16.7$ &        &        &        &         \\
\hline								  
$\QQEN$&        &        &        &        &        &        & $73.3$ &$\pz0.2$&$\pz1.6$&         \\
$\QQMN$&        &        &        &        &        &        &$\pz0.1$& $74.2$ &$\pz4.2$&         \\
$\QQTN$&        &        &        &        &        &        &$\pz6.2$&$10.1$  & $49.8$ &$\pz0.1$ \\
\hline
$\QQQQ$&        &        &        &        &        &        &$\pz0.1$&        &$\pz0.4$& $84.0$  \\
\hline
\end{tabular}
\caption[]{
  Selection efficiencies for the signal processes $\EELNLN$, $\EEQQLN$,
  and $\EEQQQQ$, at $\sqrt{s} = 206.5~\GeV$. 
  For the $\EEQQQQ$ selection, the numbers are quoted
  for a neural-network output greater than 0.6. Selection efficiencies
  at other centre-of-mass energies are only marginally different. }
\label{tab:xmat-1}
\end{center}
\end{table}

\begin{table}[p]
\begin{center}
\renewcommand{\arraystretch}{1.2}

\begin{tabular}{| c || c | c |c | c ||}
\hline
\multicolumn{1}{|c||}{$\EE \rightarrow$}  
          & $N_{\mathrm{data}}$ & $N_{\mathrm{bg}}$  
          & $\sigma(\mathrm{CC03})$ 
          & $\sigma_{\mathrm{SM}} $  \\
& & & [pb] & [pb] \\
\hline\hline
 & \multicolumn{4}{|c||}{$\langle \sqrt{s} \rangle = 188.6~\GeV$} \\
\hline
$\LNLN$ & $\pz235$ & $\pz 57.2$  &  $1.87\pm0.17\pm 0.06$  & $1.72$ \\
$\QQEN$ & $\pz347$ & $\pz 22.9$  &  $2.29\pm0.14\pm 0.03$  & $2.38$ \\
$\QQMN$ & $\pz341$ & $\pz 14.9$  &  $2.25\pm0.14\pm 0.04$  & $2.38$ \\
$\QQTN$ & $\pz413$ & $\pz  69.7$ &  $2.82\pm0.22\pm 0.07$  & $2.38$ \\
$\QQQQ$ & $ 1477$  & $    328.7$ &  $7.17\pm0.24\pm 0.12$  & $7.42$ \\
\hline \hline
& \multicolumn{4}{|c||}{$\langle \sqrt{s} \rangle = 191.6~\GeV$} \\
\hline
$\LNLN$ & $\pz 35$ & $\pz10.4$   & $1.67\pm0.41\pm 0.07$  & $1.76$ \\
$\QQEN$ & $\pz 73$ & $\pz\pz4.1$ & $2.95\pm0.37\pm 0.04$  & $2.42$ \\
$\QQMN$ & $\pz 63$ & $\pz\pz2.4$ & $2.61\pm0.36\pm 0.04$  & $2.42$ \\
$\QQTN$ & $\pz 57$ & $\pz11.9$   & $1.87\pm0.48\pm 0.05$  & $2.42$ \\
$\QQQQ$ & $  236 $ & $\pz57.5$   & $6.79\pm0.56\pm 0.15$  & $7.56$ \\
\hline \hline
& \multicolumn{4}{|c||}{$\langle \sqrt{s} \rangle= 195.5~\GeV$} \\
\hline
$\LNLN$ & $105$ & $\pz30.2$      & $1.76\pm0.25\pm 0.06$  & $1.79$ \\
$\QQEN$ & $168$ & $\pz10.9$      & $2.36\pm0.20\pm 0.03$  & $2.46$ \\
$\QQMN$ & $157$ & $\pz\pz8.2$    & $2.14\pm0.20\pm 0.03$  & $2.46$ \\
$\QQTN$ & $222$ & $\pz33.8$      & $3.44\pm0.34\pm 0.08$  & $2.46$ \\
$\QQQQ$ & $665$ & $153.5$        & $6.92\pm0.34\pm 0.11$  & $7.68$ \\
\hline \hline
& \multicolumn{4}{|c||}{$\langle \sqrt{s} \rangle = 199.6~\GeV$} \\
\hline
$\LNLN$ & $\pz87$ & $\pz26.0$      & $1.68\pm0.27\pm 0.06$  & $1.80$ \\
$\QQEN$ & $152$   & $\pz11.4$      & $2.21\pm0.20\pm 0.03$  & $2.48$ \\
$\QQMN$ & $142$   & $\pz\pz7.3$    & $2.05\pm0.20\pm 0.04$  & $2.48$ \\
$\QQTN$ & $181$   & $\pz32.2$      & $2.75\pm0.32\pm 0.07$  & $2.48$ \\
$\QQQQ$ & $726$   & $151.1$        & $7.91\pm0.36\pm 0.13$  & $7.76$ \\
\hline 
\end{tabular}\begin{tabular}{| c | c |c | c |}
\hline
          $N_{\mathrm{data}}$ & $N_{\mathrm{bg}}$ 
          & $\sigma(\mathrm{CC03})$
          & $\sigma_{\mathrm{SM}} $  \\
& & [pb] & [pb] \\
\hline\hline

\multicolumn{4}{|c|}{$\langle \sqrt{s} \rangle = 201.8~\GeV$} \\
\hline
 $\pz40$   & $\pz12.3$      & $1.47\pm0.35\pm 0.07$  & $1.81$ \\
 $\pz70$   & $\pz\pz5.3$    & $2.26\pm0.30\pm 0.03$  & $2.49$ \\
 $\pz79$   & $\pz\pz3.4$    & $2.62\pm0.33\pm 0.05$  & $2.49$ \\
 $\pz77$   & $\pz13.9$      & $2.45\pm0.47\pm 0.06$  & $2.49$ \\
 $301$     & $\pz64.6$      & $7.10\pm0.52\pm 0.12$  & $7.79$ \\
\hline\hline

\multicolumn{4}{|c|}{$\langle \sqrt{s} \rangle = 204.8~\GeV$} \\
\hline
         
 $\pz85$ & $\pz 25.9$   & $1.58\pm0.26\pm 0.05$  & $1.82$ \\
 $  176$ & $\pz 11.0$   & $2.78\pm0.23\pm 0.04$  & $2.50$ \\
 $  142$ & $\pz\pz6.5$  & $2.30\pm0.22\pm 0.04$  & $2.50$ \\
 $  164$ & $\pz  26.4$  & $2.63\pm0.33\pm 0.07$  & $2.50$ \\
 $  656$ & $    137.2$  & $7.66\pm0.37\pm 0.13$  & $7.81$ \\
\hline \hline
\multicolumn{4}{|c|}{$\langle \sqrt{s} \rangle = 206.5~\GeV$} \\
\hline
 $\pz128$ & $\pz42.6$ &  $1.42\pm0.19\pm 0.06$  & $1.82$ \\
 $\pz269$ & $\pz16.9$ &  $2.56\pm0.17\pm 0.03$  & $2.50$ \\
 $\pz240$ & $\pz11.8$ &  $2.28\pm0.17\pm 0.04$  & $2.50$ \\
 $\pz287$ & $\pz45.1$ &  $2.92\pm0.27\pm 0.07$  & $2.50$ \\
 $  1108$ & $220.1$   &  $8.07\pm0.29\pm 0.13$  & $7.82$ \\
\hline \hline
\multicolumn{4}{|c|}{$\langle \sqrt{s} \rangle = 208.0~\GeV$} \\
\hline
 $11$ & $\pz2.4$      &  $2.23\pm0.86\pm 0.06$  & $1.82$ \\
 $14$ & $\pz1.1$      &  $2.02\pm0.61\pm 0.03$  & $2.50$ \\
 $23$ & $\pz0.7$      &  $3.59\pm0.81\pm 0.05$  & $2.50$ \\
 $17$ & $\pz2.9$      &  $2.43\pm1.03\pm 0.06$  & $2.50$ \\
 $65$ & $14.1$        &  $7.28\pm1.16\pm 0.11$  & $7.82$ \\
\hline
\end{tabular}
\caption[]{
  Number of selected data events, $N_{\mathrm{data}}$, number of
  expected background events, $N_{\mathrm{bg}}$, not originating from
  W-boson pair production, and CC03 cross
  sections for the reactions $\EELNLN$, $\EEQQEN$, $\EEQQMN$, $\EEQQTN$,
  and $\EEQQQQ$. 
  For $\EEQQQQ$, $N_{\mathrm{data}}$ and $N_{\mathrm{bg}}$ correspond to a cut on the
  output of the neural network at 0.6, while the $\EEQQQQ$ cross
  section is obtained from a fit to the neural-network output
  distribution, as described in Section~\ref{sec:fitmethod}.  
  All cross sections are derived without any assumption on the W-boson
  decay branching fractions.
  The first
  uncertainty is statistical and the second systematic.  Also shown are the
  Standard Model  CC03 cross sections, $\sigma_{\mathrm{SM}}$,
  as calculated with YFSWW3~\cite{YFSWW3} with an 
  uncertainty of 0.5\%.
\label{tab:xsec1}
}
\end{center}
\end{table}

\begin{table}[p]
\begin{center}
\renewcommand{\arraystretch}{1.2}
\begin{tabular}{|c|cc|}
\hline
$\langle \sqrt{s} \rangle \, [\GeV]$ & $\sigma^\mathrm{meas}_{\mathrm{qq}}\; [\mathrm{pb}]$ & 
$\sigma^\mathrm{MC}_{\mathrm{qq}}\; [\mathrm{pb}]$  \\
\hline
188.6 & $107.5\pm3.4$      &  $101.00$   \\         
191.6 & $\pz92.8\pm7.6$    &  $\pz97.74$ \\  	       
195.5 & $\pz86.7\pm4.5$    &  $\pz92.47$ \\  	         
199.6 & $\pz86.8\pm4.7$    &  $\pz88.09$ \\       
201.8 & $\pz89.6\pm7.0$    &  $\pz85.89$ \\  	         
204.8 & $\pz84.1\pm4.7$    &  $\pz82.19$ \\       
206.5 &	$\pz78.1\pm3.6$    &  $\pz80.90$ \\
\hline
\end{tabular}
\caption[]{
Measured, $\sigma^\mathrm{meas}_{\mathrm{qq}}$,  and expected,
$\sigma^\mathrm{MC}_{\mathrm{qq}}$,  
cross sections of the $\EEQQG$ process.
The measurements are determined by a fit of the neural-network output distribution of the $\QQQQ$ 
selection to both signal and $\EEQQG$ background.}
\label{tab:xsecqq}
\end{center}
\end{table}

\begin{table}[p]
\begin{center}
\renewcommand{\arraystretch}{1.2}
\newcommand{\phm}{\phantom{$<$}}
\begin{tabular}{|l||c|c|c|c|c|}
\hline
\multicolumn{6}{|c|}{Systematic uncertainties on $\sigma~[\%]$}\\
\hline
& \multicolumn{5}{|c|}{Final state} \\
\cline{2-6}
Source                      &$\LNLN$&$\QQEN$&$\QQMN$&$\QQTN$&$\QQQQ$\\
\hline
Luminosity                  & \multicolumn{5}{c|}{0.22}        \\
\cline{2-6}
MC statistics (signal)      &   0.80 &  0.25 &  0.25 &  0.44 & \phm 0.11 \\
MC statistics (background)  &   1.57 &  0.23 &  0.28 &  0.75 & \phm 0.22 \\
Detector modelling          &   2.00 &  1.00 &  1.20 &  2.00 & \phm 1.00 \\
Hadronisation (signal)      &   ---  &  0.77 &  0.58 &  1.17 & \phm 0.45 \\ 
Hadronisation (background)  &   ---  &  ---  &  ---  &  ---  & \phm 0.90 \\ 
\cline{3-5}
Bose-Einstein effects       &   --- &  \multicolumn{3}{c|}{$<0.01$}& \phm 0.03 \\ 
\cline{3-5}
Colour reconnection         &   --- &  --- &  --- &  --- &  \phm 0.19 \\ 
Background cross sections   &   0.59 &  0.21 &  0.22 &  0.40 & \phm 0.40 \\
W mass  ($\pm0.04~\GeV$)    &   0.27 &  0.03 &  0.03 &  0.10 & \phm 0.06 \\ 
W width ($\pm0.06~\GeV$)    &   0.12 &  0.03 &  0.12 &  0.08 & \phm 0.02 \\
\cline{2-6}
ISR simulation              &   \multicolumn{5}{c|}{$<0.01$} \\
\cline{2-6}
FSR simulation              &   0.21 &  0.21 &  0.17 &  0.08 & $<0.01$  \\ 
\hline
Total                       &   2.76 &  1.36 &  1.43 &  2.52 &  \phm 1.46 \\
\hline
\end{tabular}
\caption[]{
  Relative systematic uncertainties on the cross-section
  measurements evaluated for $\sqrt{s}=206.5~\GeV$. Uncertainties at
  other center-of-mass energies are only marginally different.  
\label{tab:syst}
}
\end{center}
\end{table}

\begin{table}[p]
\begin{center}
\renewcommand{\arraystretch}{1.2}
\begin{tabular}{| c || c | c | c || c |}
\hline
$\sigma$ [pb] & 
$\langle \sqrt{s} \rangle = 188.6$ $\GeV$ &
$\langle \sqrt{s} \rangle = 191.6$ $\GeV$ &
$\langle \sqrt{s} \rangle = 195.5$ $\GeV$ &
$\langle \sqrt{s} \rangle = 199.6$ $\GeV$ \\
\hline
$\sigma_{\LNLN}$ & 
$1.88\pm0.16\pm0.07$ & $1.66\pm0.39\pm0.07$ & $1.78\pm0.24\pm0.07$ & $1.75\pm0.25\pm0.06$ \\
$\sigma_{\QQLN}$  &
$7.19\pm0.24\pm0.08$ & $7.69\pm0.61\pm0.09$ & $7.58\pm0.36\pm0.08$ & $6.81\pm0.35\pm0.08$ \\
$\sigma_{\QQQQ}$  &
$7.17\pm0.24\pm0.12$ & $6.78\pm0.56\pm0.12$ & $6.92\pm0.34\pm0.11$ & $7.91\pm0.36\pm0.13$ \\
\hline
$\sigma_{\mathrm{WW}}$  &
$16.17\pm0.37\pm0.17$ & $16.11\pm0.89\pm0.17$ & $16.22\pm0.54\pm0.16$ & $16.49\pm0.55\pm0.17$ \\
\hline
$\sigma_{\mathrm{SM}}$  &
$16.27$ & $16.57$ & $16.84$ & $17.02$ \\
\hline\hline
$\sigma$ [pb] & 
$\langle \sqrt{s} \rangle = 201.8$ $\GeV$ &
$\langle \sqrt{s} \rangle = 204.8$ $\GeV$ &
$\langle \sqrt{s} \rangle = 206.5$ $\GeV$ &
$\langle \sqrt{s} \rangle = 208.0$ $\GeV$ \\
\hline
$\sigma_{\LNLN}$  & 
$1.51\pm0.34\pm0.07$ & $1.58\pm0.24\pm0.05$ & $1.44\pm0.18\pm0.06$ & $2.23\pm0.86\pm0.06$ \\
$\sigma_{\QQLN}$  &
$7.34\pm0.54\pm0.08$ & $7.68\pm0.39\pm0.13$ & $7.60\pm0.30\pm0.08$ & $8.18\pm1.21\pm0.09$ \\
$\sigma_{\QQQQ}$  &
$7.09\pm0.52\pm0.12$ & $7.66\pm0.37\pm0.13$ & $8.07\pm0.29\pm0.13$ & $7.29\pm1.16\pm0.11$ \\
\hline
$\sigma_{\mathrm{WW}}$  &
$16.01\pm0.81\pm0.17$ & $17.00\pm0.58\pm0.17$ & $17.31\pm0.45\pm0.18$ & $17.52\pm1.81\pm0.17$ \\
\hline
$\sigma_{\mathrm{SM}}$  &
$17.08$ & $17.12$ & $17.14$ & $17.15$ \\
\hline
\end{tabular}
\vskip 0.5cm

\caption[]{Measured CC03 cross sections of the processes
  $\EELNLN$, $\EEQQLN$ (summed over lepton flavours) and  $\EEQQQQ$,
assuming charged-lepton universality.
  The measured W-boson pair-production cross sections, $\mathrm{\sigma_{WW}}$, are derived 
  assuming Standard Model branching fractions for the W boson decay
  modes. The Standard Model total W-boson pair-production  cross
  sections, $\sigma_{\mathrm{SM}}$, are calculated
  using the YFSWW3 program, which has a theoretical uncertainty 
  of 0.5\%.
\label{tab:sigma3}
}
\end{center}
\end{table}

\begin{table}[p]
\begin{center}
\renewcommand{\arraystretch}{1.2}
\begin{tabular}{| c || c | c || c |}
\hline
Branching &          Lepton             &    Lepton             & Standard \\
fraction  &        non-universality    &  universality         & Model    \\
\hline\hline
$\mathrm{Br}(\WEN)~[\%]$& $  10.78\pm0.29\pm0.13$ &      ---              &      \\
$\mathrm{Br}(\WMN)~[\%]$& $  10.03\pm0.29\pm0.12$ &      ---              &      \\
$\mathrm{Br}(\WTN)~[\%]$& $  11.89\pm0.40\pm0.20$ &      ---              &      \\
$\mathrm{Br}(\WLN)~[\%]$&        ---              & $10.83\pm0.14\pm0.10$ & $10.83$ \\
$\mathrm{Br}(\WQQ)~[\%]$& $  67.30\pm0.42\pm0.30$ & $67.50\pm0.42\pm0.30$ & $67.51$ \\
\hline
\end{tabular}
\vskip 0.5cm
\caption[]{W-boson decay branching fractions 
  derived without and with the
  assumption of charged-lepton universality.  The correlation coefficients
  between the leptonic branching fractions are $-0.016$, $-0.279$,
  $-0.295$ for [$\mathrm{Br}(\WEN), \mathrm{Br}(\WMN)$],
  [$\mathrm{Br}(\WEN),\mathrm{Br}(\WTN)$] and 
  [$\mathrm{Br}(\WMN),\mathrm{Br}(\WTN)$], respectively.  
  The W-boson decay branching fractions
  expected in the Standard Model are also listed. 
\label{tab:brafra}
}
\end{center}
\end{table}


\begin{table}[p]
\begin{center}
\renewcommand{\arraystretch}{1.2}
\begin{tabular}{| c || c | c || c | c |}
\cline{2-5}
\multicolumn{1}{c}{} &
\multicolumn{4}{|c|}{$\mathrm{d}\sigma/\mathrm{d}\ctw$ [pb]} \\
\hline
$\cos\theta_\mathrm{W^-}$ range &$\langle \sqrt{s} \rangle = 182.7$ & SM &
$\langle \sqrt{s} \rangle = 189.0$ & SM \\
\hline
$-1.0,-0.8$       & $0.54\pm0.23\pm0.01$ & 0.74 &$0.69\pm0.12\pm0.01$ & 0.64  \\
$-0.8,-0.6$       & $0.81\pm0.29\pm0.01$ & 0.84 &$0.88\pm0.15\pm0.01$ & 0.78  \\
$-0.6,-0.4$       & $0.22\pm0.26\pm0.00$ & 1.02 &$1.08\pm0.17\pm0.02$ & 0.94  \\
$-0.4,-0.2$       & $0.96\pm0.33\pm0.01$ & 1.20 &$1.18\pm0.19\pm0.02$ & 1.14  \\
$-0.2,\pmi0.0$    & $1.71\pm0.43\pm0.03$ & 1.44 &$1.34\pm0.20\pm0.02$ & 1.38  \\
$\pmi0.0,\pmi0.2$ & $2.27\pm0.50\pm0.03$ & 1.78 &$1.51\pm0.22\pm0.02$ & 1.72  \\
$\pmi0.2,\pmi0.4$ & $3.37\pm0.62\pm0.05$ & 2.16 &$1.88\pm0.24\pm0.03$ & 2.22  \\
$\pmi0.4,\pmi0.6$ & $3.52\pm0.66\pm0.05$ & 2.86 &$2.95\pm0.31\pm0.04$ & 2.95  \\
$\pmi0.6,\pmi0.8$ & $4.24\pm0.74\pm0.06$ & 3.84 &$4.19\pm0.37\pm0.06$ & 4.15  \\
$\pmi0.8,\pmi1.0$ & $5.00\pm0.83\pm0.07$ & 5.47 &$6.11\pm0.47\pm0.09$ & 6.24  \\
\hline
\end{tabular}

\vskip 1cm

\begin{tabular}{| c || c | c || c | c |}
\cline{2-5}
\multicolumn{1}{c}{} &
\multicolumn{4}{|c|}{$\mathrm{d}\sigma/\mathrm{d}\ctw$ [pb]} \\
\hline
$\cos\theta_\mathrm{W^-}$ range &$\langle \sqrt{s} \rangle = 198.3$ & SM &
$\langle \sqrt{s} \rangle = 205.9$ & SM \\
\hline
$-1.0,-0.8$       & $0.68\pm0.11\pm0.01$ & 0.57 &$0.60\pm0.10\pm0.01$ &0.52   \\
$-0.8,-0.6$       & $0.76\pm0.13\pm0.01$ & 0.71 &$0.44\pm0.11\pm0.01$ &0.64   \\
$-0.6,-0.4$       & $0.78\pm0.15\pm0.01$ & 0.85 &$0.77\pm0.14\pm0.01$ &0.78   \\
$-0.4,-0.2$       & $0.80\pm0.16\pm0.01$ & 1.05 &$0.99\pm0.16\pm0.01$ &0.98   \\
$-0.2,\pmi0.0$    & $1.31\pm0.20\pm0.02$ & 1.29 &$1.35\pm0.20\pm0.02$ &1.21   \\
$\pmi0.0,\pmi0.2$ & $1.64\pm0.23\pm0.02$ & 1.65 &$1.72\pm0.23\pm0.03$ &1.55   \\
$\pmi0.2,\pmi0.4$ & $2.21\pm0.27\pm0.03$ & 2.16 &$1.75\pm0.23\pm0.03$ &2.06   \\
$\pmi0.4,\pmi0.6$ & $2.41\pm0.29\pm0.04$ & 2.97 &$2.84\pm0.30\pm0.04$ &2.92   \\
$\pmi0.6,\pmi0.8$ & $3.69\pm0.36\pm0.05$ & 4.38 &$4.80\pm0.41\pm0.07$ &4.45   \\
$\pmi0.8,\pmi1.0$ & $6.26\pm0.49\pm0.09$ & 7.20 &$7.49\pm0.53\pm0.11$ &7.80   \\
\hline											
\end{tabular}

\vskip 0.5cm

\caption[]{Sum of the differential cross sections, as function of
  $\ctw$, for the $\EEQQEN$ and $\EEQQMN$ processes. 
The measurements are derived in a restricted phase space of the CC03
  subset of diagrams.
The first uncertainty is statistical and the second is systematic.
The systematic uncertainty is fully correlated between 
$\cos\theta_\mathrm{W^-}$ bins and between $\sqrt{s}$ bins.
The columns labeled SM show the expected
values from the Standard Model, which have a theoretical uncertainty
of about 2\%.
\label{tab:diffxs}
}
\end{center}
\end{table}

\clearpage

\clearpage

\clearpage

\end{document}